\documentclass[12pt]{article}

\usepackage{graphicx}
\usepackage{jheppub}

\usepackage{amsmath,amssymb,array,mathrsfs,amsfonts}
\usepackage{epsfig}
\usepackage{euscript}
\usepackage{subfigure}

\usepackage{bbm}

\newcommand{\beq}{\begin{equation}}
\newcommand{\eeq}{\end{equation}}
\newcommand{\bea}{\begin{eqnarray}}
\newcommand{\eea}{\end{eqnarray}}
\renewcommand{\th}{\theta}
\newcommand{\rf}[1]{(\ref{#1})}
\newcommand{\ra}{\rightarrow}
\newcommand{\pa}{\partial}
\newcommand{\oh}{\frac{1}{2}}
\newcommand{\oq}{\frac{1}{4}}
\newcommand{\non}{\nonumber}
\newcommand*{\Dsl}[0]{{\rlap{\kern2.25pt /}{D}}}
\newcommand*{\Asl}[0]{{\rlap{\kern2.25pt /}{A}}}
\newcommand*{\dsl}[0]{{\rlap{\kern0.5pt /}{\partial}}}
\newcommand*{\xisl}[0]{{\rlap{\kern0.5pt /}{\xi}}}
\newcommand*{\asl}[0]{{\rlap{\kern0.5pt /}{a}}}
\newcommand*{\bsl}[0]{{\rlap{\kern0.5pt /}{b}}}

\newcommand*{\tr}[0]{{\rm tr}}

\def\Dslash{\,\,{\raise.15ex\hbox{/}\mkern-12mu D}}

\newcommand{\SP}[1]{\begin{equation}\begin{split} #1
\end{split}\end{equation}}

\def\b{\beta}

\def\d{\delta}

\def\m{\mu}

\def\r{\rho}
\def\s{\sigma}

\def\D{\Delta}

\newcommand{\Tr}{\operatorname{Tr}}

\def\B0{{\boldsymbol 0}}

\def\tr{{\rm tr}}

\def\Tr{{\rm Tr}}

\def\det{{\rm det}}

\def\Dbarslash{\,\,{\raise.15ex\hbox{/}\mkern-12mu {\bar D}}}
\def\Dslash{\,\,{\raise.15ex\hbox{/}\mkern-12mu D}}
\def\delslash{\,\,{\raise.15ex\hbox{/}\mkern-9mu \partial}}
\def\delbarslash{\,\,{\raise.15ex\hbox{/}\mkern-9mu {\bar\partial}}}

\newcommand\nn{\nonumber}

\newcommand{\EQ}[1]{\begin{equation}\begin{split} #1
\end{split}\end{equation}}

\title{The density in the density of states method}
\author[a,b]{Jeff Greensite}
\author[c]{Joyce C. Myers}
\author[c]{and K. Splittorff}
\affiliation[a]{Physics and Astronomy Department, San Francisco State University, 1600 Holloway Ave., San Francisco CA, 94132 USA}
\affiliation[b]{Niels Bohr International Academy, Blegdamsvej 17, 2100 Copenhagen {\O}, Denmark}
\affiliation[c]{Discovery Centre, The Niels Bohr Institute, University of Copenhagen, Blegdamsvej 17, 2100 Copenhagen {\O}, Denmark}
\emailAdd{greensit@sfsu.edu, jcmyers@nbi.dk, split@nbi.dk}
\abstract{
It has been suggested that for QCD at finite baryon density the distribution of the phase angle, i.e.\ the angle defined as the imaginary part of the logarithm of the fermion determinant, has a simple Gaussian form.  This distribution provides the density in the density of states approach to the sign problem. We calculate this phase angle distribution using i) the hadron resonance gas model; and ii) a combined strong coupling and hopping parameter expansion in lattice gauge theory.  While the former model leads only to a Gaussian distribution, in the latter expansion we discover terms which cause the phase angle
distribution to deviate, by relative amounts proportional to powers of the inverse lattice volume, from a simple Gaussian form. We show that despite the tiny inverse-volume deviation of the phase angle distribution from a simple Gaussian form, such non-Gaussian terms can have a substantial impact on observables computed in the density of states/reweighting approach to the sign problem.
}

\notoc
\begin{document}

\bibliographystyle{h-physrev5}

\maketitle

\newpage


\section{Introduction}

   The QCD action at finite chemical potential is complex, and therefore the usual technique of numerical simulation via importance sampling is not directly applicable. This is the ``sign problem,'' see 
\cite{Aarts:2013bla,deForcrand:2010ys,Hands:2007by,Splittorff:2006vj} for some recent reviews. Since the difficulty originates in the complex phase $e^{i\th}$ of the fermion determinant, where
\begin{equation}
   \det(\Dsl+\gamma_0 \mu + m) = |\det(\Dsl+\gamma_0 \mu + m)| e^{i \theta} \ ,
\end{equation}
an obvious strategy is to treat the complex phase factor $e^{i\th}$ as an operator, rather than as part of the action.  This general approach to the sign problem is known as ``reweighting."  Suppose, for example, that one would like to evaluate the expectation value of some operator $X$ in a theory with $N_f$ flavors of fermions with equal masses.  Define the partition function $Z$ and the ``phase-quenched'' partition function $Z_{pq}$ as
\begin{eqnarray}
     Z &=& \int DU |\det^{N_f}(\Dsl+\gamma_0 \mu + m)| e^{i N_f \theta} e^{-S_{YM}}
\nonumber \\
     Z_{pq} &=& \int DU |\det^{N_f}(\Dsl+\gamma_0 \mu + m)| e^{-S_{YM}}
\label{Dirac}
\end{eqnarray}
with corresponding expectation values in the full and phase-quenched distributions denoted $\langle...\rangle_{QCD}$ and
$\langle...\rangle_{pq}$ respectively.  $S_{YM}$ is the action of the pure gauge theory, and the notation 
$\langle...\rangle$ denotes expectation values in that pure gauge theory, with no matter fields in the integration measure.  Then, if $X$ denotes some observable, we have
\begin{eqnarray}
          \langle X \rangle_{QCD} = { \langle X e^{i N_f \theta} \rangle_{pq} \over  \langle  e^{i N_f \theta} \rangle_{pq} } \ ,
\label{ratio}
\end{eqnarray}
where both the numerator and denominator on the right-hand side could, in principle, be evaluated via importance sampling.
The catch is that when the sign problem is severe, both the numerator and denominator of \rf{ratio} are 
extraordinarily small, e.g.\ \cite{Splittorff:2006fu,Splittorff:2007zh}
\bea
 \langle  e^{i N_f \theta} \rangle_{pq} &=& { Z \over Z_{pq}}  = \exp[ -\D F ]
\non \\
     &=& \exp[-c V] \ ,
\label{very_small}
\eea
where $\D F$ is the difference in free energies of the full and phase-quenched theories, $V$ is the lattice 3-volume (i.e.\ the spatial volume of a time slice), and $c$ is some volume-independent constant.  In most cases of interest eq.\ \rf{very_small} implies a value of 
$\langle  e^{i N_f \theta} \rangle_{pq}$ which is so small as to require, from importance sampling, a level of accuracy which is simply unattainable in practice.

   One may introduce a probability distribution $\rho_{pq}(\th)$ for the phase angle of the fermion determinant in the phase quenched theory, such that for any function ${\cal O}(\th)$,
\beq
         \langle {\cal O} \rangle_{pq} = \int d\th ~ {\cal O}(\th) \rho_{pq}(\th)  \ .
\label{angvev}
\eeq
This is essential to the ``density of states'' approach \cite{Gocksch:1988iz,Ejiri:2007ga,Fukushima:2010bq,Anagnostopoulos:2001yb,Fodor:2007vv}, which is closely related to reweighting.  But one would still have to know the probability distribution (or ``density'') $\rho_{pq}(\th)$ to fantastic accuracy, in order to overcome the exponential volume suppression and extract the correct value of $\langle  e^{i N_f \theta} \rangle_{pq}$.
  
    Recently Ejiri and the WHOT-QCD collaboration \cite{Ejiri:2007ga,Nakagawa:2011eu,Ejiri:2012ng,Ejiri:2012wp,Ejiri:2013lia},  have developed a method to circumvent this 
$e^{-cV}$ signal suppression, which is the essence of the sign problem in the reweighting/density of states approach.  Their method relies on
\begin{itemize}
\item the conjecture that the distribution of the phase angle $\th$ has a simple Gaussian form;
\item the direct measurement of this distribution by a histogram technique;
\item the evaluation of $\langle  e^{i N_f \theta} \rangle_{pq}$ (or some variant thereof) by the method of cumulants;
\item the fact that the conjectured Gaussian form of the phase angle distribution implies that only the second (and easiest to measure) cumulant is non-zero.
\end{itemize}
This approach, along with some other innovations in refs.\ \cite{Ejiri:2007ga,Ejiri:2012ng,Ejiri:2012wp,Ejiri:2013lia}, amounts to a very sophisticated version of the reweighting approach, and leads to an effective action
subsequently used for determining the phase diagram of QCD.

    In this article we will critically examine certain aspects of this improved version of reweighting, in particular the arguments which lead to the conclusion that only the second cumulant is required in the cumulant expansion of the complex phase.  Our study will suggest that although the phase angle distribution derived from the histogram technique may \emph{appear} almost indistinguishable from a Gaussian, tiny corrections to that Gaussian may result in significant higher-order cumulants, which cannot be neglected. Evaluation of such cumulants will require measurement of moments of the phase angle to a relative accuracy of at least one power, and quite possibly higher powers, of the inverse lattice volume. 
    
    In section \ref{pad} below we will justify the statements made in the last paragraph.  We then show in detail our computation of the phase angle distribution in the hadron resonance gas model (section \ref{hrg}), and in lattice QCD via the 
strong coupling/hopping parameter expansion (section \ref{sc}).  In section \ref{taylor} we present an argument for the
existence of non-Gaussian contributions which is independent of strong-coupling and hopping parameter expansions. Our conclusions are found in section \ref{conclude}. Further details are reserved for the appendices.
 
\section{\label{pad} Cumulants and the Phase Angle Distribution} 

   Let us define the normalized phase-angle distributions in the full and phase-quenched theories
\bea\label{rhodef}
        \rho(\th') &=& {1\over Z} \int DU \d[\th'-\th(U)] \det^{N_f}(\Dsl+\gamma_0 \mu + m) e^{-S_g}
\non \\
          \rho_{pq} (\th') &=& {1\over Z_{pq}} \int DU \d[\th'-\th(U)] \Bigl|\det^{N_f}(\Dsl+\gamma_0 \mu + m)\Bigr| e^{-S_g} \ ,
\eea
where $\th(U)$ is the phase of the fermion determinant defined in \rf{Dirac}.  By inspection \cite{Splittorff:2007zh}
\beq
   \rho(\th) = {Z_{pq} \over Z} e^{iN_f \th}  \rho_{pq} (\th) \ .
\label{rho_full_pq}
\eeq
We may also define restricted distributions
\beq
\rho_{pq} (\th',X') = {1\over Z_{pq}} \int DU \d[\th'-\th(U)] \d[X'-X(U)] |\det(\Dsl+\gamma_0 \mu + m)| e^{-S_g}  \ ,
\eeq
where $X$ is some observable or set of observables such as the average Polyakov line, plaquette energy, or magnitude
of the fermion determinant.  A distribution of this kind could be used to compute expectation values via reweighting, i.e.
\beq
\langle X \rangle_{QCD} = {\int dX d\th ~ X e^{i N_f \th} \rho_{pq} (\th,X) \over \int d\th ~ e^{i N_f \th}  \rho_{pq} (\th) } \ ,
\eeq
or to extract an effective potential as outlined in \cite{Ejiri:2013lia},  but here we will be mainly concerned with the phase angle distribution $\rho_{pq} (\th)$ in the denominator.  Given this distribution, we have
\beq
\langle e^{iN_f \th} \rangle_{pq} = \int d\th ~  e^{iN_f \th} \rho_{pq} (\th)  \  .
\eeq
The distribution  $\rho_{pq} (\th)$ can be computed by a histogram approach, i.e.\ generating gauge configurations via importance sampling in the phase-quenched measure, counting the number of times that the phase angle $\th$ falls into each of a large set of intervals, and finally normalizing the result.   But at this point we confront the sign problem:  the fact that
$e^{iN_f \th}$ is an oscillating function whose positive and negative contributions cancel \emph{almost} exactly, leaving
a tiny net value of order  $e^{-cV}$.  This means that 
$\rho_{pq} (\th)$ would have to be computed to fantastic accuracy in order to produce such a nearly exact cancellation. The interesting suggestion of \cite{Ejiri:2007ga,Ejiri:2012ng,Ejiri:2012wp,Ejiri:2013lia} is to deal with this problem by the method of cumulants, i.e.
\beq
    \langle e^{iN_f \th} \rangle_{pq} = \exp\left[ \sum_{n=1}^\infty  {(-N_f^2)^n \over (2n)!} (\th^{2n})_c \right]  \ ,
\label{cumulant_expansion}
\eeq    
where the first few cumulants are given by
\bea
(\th^2)_c &=& \langle \th^2 \rangle_{pq}
\non \\
(\th^4)_c &=& \langle \th^4 \rangle_{pq} - 3 \langle \th^2 \rangle_{pq}^2
\non \\
(\th^6)_c &=& \langle \th^6 \rangle_{pq} - 15 \langle \th^4 \rangle_{pq}\langle \th^2 \rangle_{pq} 
+ 30 \langle \th^2 \rangle_{pq}^3  \ .
\label{cumulants}
\eea
Even powers of $\th$ are monotonically increasing rather than oscillating, and the cumulants are all of $O(V)$, which immediately leads to  $ \langle e^{iN_f \th} \rangle_{pq}\sim e^{-cV}$.   On the other hand, since the higher moments are proportional to higher powers of $V$, i.e.\
$\langle \th^{2n} \rangle \sim V^n$ (as we show in the next subsection), 
it is still necessary to compute moments to a very high degree of accuracy,
in order that 4th and higher-order cumulants, which involve delicate cancellations among moments, will come out to depend only linearly on the volume.  For example, $(\th^4)_c$ is $O(V)$, yet the moments $\langle \th^4 \rangle_{pq}$ and
$\langle \th^2 \rangle^2_{pq}$ are $O(V^2)$.  So to get the proper cancellation of the $O(V^2)$ pieces, leaving an
$O(V)$ remainder for the cumulant, it is necessary to 
measure the moments to a relative accuracy of $1/V$.  Higher cumulants will require relative accuracy of the moments to still higher powers of the inverse volume.    For this reason, it is important to know whether the phase angle distribution has a purely Gaussian form,
as conjectured in \cite{Ejiri:2007ga,Splittorff:2007zh,Lombardo:2009aw}, in which case only the 2nd order cumulant 
$(\th^2 )_c$ is non-zero, or whether there are significant corrections to the Gaussian approximation, which means that higher order cumulants will also have to be computed.

    At this point one might object that the phase of the fermion determinant lies between $-\pi$ and $\pi$, not between $-\infty$
and $\infty$.  However, in the case of QCD the range of the phase angle of the fermion determinant can also be taken to be infinite, rather than a finite interval of length $2\pi$.  It is mainly a matter of how the phase angle is actually extracted. 
Let ${\cal M}=\Dsl+\gamma_0 \mu + m$, and $T$ denote temperature.  The unrestricted range comes about in a natural way when the phase angle is determined from the \\ formula \cite{Nakagawa:2011eu} 
\bea
\th &\equiv& \mbox{Im}\log(\det {\cal M}) 
\non \\
&=&  \int_0^{\m/T} \mbox{Im}\left[{\pa (\log \det  {\cal M}) \over \pa \m/T} \right]_{\m=\overline{\m}}d\left({\overline{\m} \over T}\right)
\non \\
&=&  \int_0^{\m/T} \tr\left[{\cal M}^{-1} {\pa {\cal M} \over \pa \m/T} \right]_{\m=\overline{\m}}d\left({\overline{\m} \over T}\right)  \ ,
\label{extract}
\eea
which is how this angle is extracted in simulations.   It is this expression for $\th$ (third line of \rf{extract}) that we have in mind, when we discuss the phase angle distribution.  However, our overall conclusions would remain the same whether we work in a finite or infinite interval. All that changes if we work with $\theta$ in $[-\pi,\pi]$ is that the $\delta$-function of Eq.~(\ref{rhodef}) is periodic as in \cite{Greensite:2013vza}.

    The question of whether the phase angle distribution is Gaussian can be investigated in a number of ways, and there does appear to be strong evidence in support of this conjecture.  For one thing, the Gaussian distribution is argued to be a natural consequence of the Central Limit Theorem \cite{Ejiri:2007ga}, and a one-loop calculation from chiral perturbation theory indicates that a free gas of pions does indeed generate a Gaussian distribution for $\rho_{pq}(\th)$ \cite{Splittorff:2007zh,Lombardo:2009aw} provided that $\mu<m_\pi/2$.
More concretely, the phase angle distribution has been evaluated by a histogram technique 
\cite{Nakagawa:2011eu,Ejiri:2007ga}, and it appears (at least by eye) to be indistinguishable from a Gaussian distribution.  The Binder cumulant
\beq
            B_4 = {\langle \th^4 \rangle_{pq} \over \langle \th^2 \rangle_{pq}^2}
\label{b4}
\eeq
has also been computed numerically \cite{Ejiri:2007ga}, and for most parameter values it seems to be perfectly consistent, within statistical error, with $B_4 =3$.  The value $B_4=3$ implies that the 4th order cumulant $(\th^4)_c$ vanishes, as expected for a Gaussian distribution in which all cumulants $(\th^{2n})_c$ vanish for $n>1$.    Despite all this, we do not believe that the existing evidence has yet ruled out the existence of significant higher-order (and hard to measure) cumulants, for reasons that we will now explain. \\

\subsection{Fourier components of the phase angle distribution}    

   Since $\rho(\th)$ is the expectation value of a delta function, we may write
\bea
  \rho(\th') &=& \langle \d[\th'-\th(U)] \rangle_{QCD}
\non \\
&=& 2 \int_{-\infty}^\infty {dp \over 2\pi} e^{-2ip\th'} \langle e^{2ip\th} \rangle_{QCD}      \ .
\label{rho}
\eea
The moments $\langle e^{2ip\th} \rangle_{QCD}$ can be expressed in the form
\beq
\langle e^{2 i p \theta} \rangle_{QCD} = \frac{Z_{YM}}{Z} \bigg{\langle} \frac{\det^p (\Dsl+\gamma_0 \mu + m)}{\det^p(\Dsl-\gamma_0 \mu + m)} {\det}^{N_f} (\Dsl + \gamma_0 \mu + m) \bigg{\rangle} \, ,
\label{complex-phase}
\eeq
where $Z_{YM}$ is the partition function of the gauge theory with no matter fields, and $\langle ... \rangle$ denotes the corresponding expectation value with a pure Yang-Mills probability measure.  In this form, the moments can be computed analytically in certain models and in certain limits of full QCD.  In both the hadron resonance gas model and in the limit of strong gauge coupling and large quark masses, it can be shown that the moments have the form
\bea
\langle e^{2ip\th} \rangle_{QCD} = \exp[-p(p+N_f)X_1 - p^2(p+N_f)^2 X_2- p^3(p+N_f)^3 X_3 - ...]  \ ,
\label{form}
\eea 
where the $X_n$ are all proportional to volume.  This volume dependence is clear because the expectation value in 
\rf{complex-phase}
has the interpretation of a partition function of $p+N_f$ quarks with chemical potential $\m$, and $p$ ``ghost quarks," i.e.\ fermions which commute rather than anticommute, having chemical potential $-\m$.  Note that the form \rf{form} respects the invariances of the moments in \rf{complex-phase}, see \cite{Greensite:2013vza} for details.  It will be the task of later sections to actually derive the $X_n$
in the hadron resonance gas model and via a strong-coupling/hopping parameter expansion of lattice QCD.  In the latter expansion, it will be shown explicitly that the $X_{n>1}$ are non-zero, and that they increase as the gauge coupling $\b$, the
hopping parameter $h$, and the chemical potential $\m$ increase.  The details will be presented in section \ref{sc}.
For now we ask the reader to provisionally accept the statement that the $X_{n>1}$ are non-zero, and allow us to explore the consequences.  

    From \rf{rho} and \rf{form}, and a change of variables $q=p-\oh N_f$, we have
\bea
\rho(\th) &=& \int {dp \over \pi} e^{-2ip\th} \exp\left[-\sum_{n=1}^\infty [p(p+N_f)]^n X_n \right]
\non \\
&=& \int {dq \over \pi} e^{-2i(q-\oh N_f)\th} \exp\left[-\sum_{n=1}^\infty [q^2 - \oq N_f^2]^n X_n \right]
\non \\
&=& e^{i N_f \th} \exp\left[-\sum_{n=1}^\infty (- \oq N_f^2)^n X_n \right] \int {dq \over \pi} e^{-2iq\th} \exp\left[-\sum_{n=1}^\infty 
q^{2n} Y_n \right]  \ ,
\label{rho_full}
\eea
where
\bea
     Y_1 &=& X_1 - \oh N_f^2  X_2 + {3\over 16} N_f^4 X_3 - {1\over 16} N_f^6 X_4 + ...
\non \\
     Y_2 &=& X_2 - {3\over 4} N_f^2 X_3 + {3\over 8} N_f^4 X_4 - ...
\non \\
     Y_3 &=& X_3 - N_f^2 X_4 + ..
\non \\
     Y_4 &=& X_4 - ...  \ .
\eea
To relate the phase angle distribution $\rho(\th)$ of the full theory to the distribution $\rho_{pq}(\th)$ relevant to reweighting, we multiply both sides of \rf{rho_full_pq} by $e^{-iN_f \th}$ and integrate over $\th$, which gives us
\bea
  \langle e^{-iN_f \th} \rangle_{QCD} = {Z_{pq} \over Z}  \ .
\eea
But we also have
\bea
{Z_{pq} \over Z} &=&  \langle e^{-iN_f \th} \rangle_{QCD} 
\non \\
&=& \int d\th e^{-iN_f \th} \rho(\th)
\non \\
&=& \int d\th   \exp\left[-\sum_{n=1}^\infty (- \oq N_f^2)^n X_n \right] \int {dq \over \pi} e^{-2iq\th} \exp\left[-\sum_{n=1}^\infty 
q^{2n} Y_n \right]
\non \\
&=&  \exp\left[-\sum_{n=1}^\infty (- \oq N_f^2)^n X_n \right]  \ .
\eea
Inserting this result for $Z_{pq}/Z$ into \rf{rho_full_pq}, and using \rf{rho_full}, we finally obtain the angular distribution for the phase-quenched theory
\bea
\rho_{pq}(\th) = \int {dq \over \pi} e^{-2iq\th}  \exp\left[-\sum_{n=1}^\infty 
q^{2n} Y_n \right] \ .
\label{rhopq}
\eea
From this expression we can easily derive the moments $\langle \th^{2n}\rangle_{pq}$ and the cumulants $(\th^{2n})_c$
in terms of the $Y_n$.  The moments are given by
\beq
\langle \th^n \rangle_{pq} = \bigg\{\left({1\over 2i}\right)^n {d^n \over dq^n} \exp\left[ -\sum_{m=1}^\infty q^{2m} Y_m \right]\bigg\}_{q=0}  \ .
\eeq
Since the $Y_n \propto V$ are extensive, it is not hard to see that $\langle \th^{2n} \rangle_{pq}$ is of order $V^n$.
We also have, from \rf{angvev} and \rf{rhopq}
\beq
\langle e^{2 i q \theta} \rangle_{pq} = e^{-q^2 Y_1 - q^4 Y_2 - ...}  \ ,
\eeq
and the cumulant expansion is defined, for arbitrary $p$,
\beq
\ln \langle e^{2 i p \theta} \rangle_{pq} = \sum_{n=1}^{\infty} \frac{(2 i p)^n}{n!}  (\theta^n )_c  \ .
\eeq
Then equating
\beq
-q^2 Y_1 - q^4 Y_2 - ... = \sum_{n=1}^{\infty} \frac{(2 i q)^n}{n!}  (\theta^n )_c  \ ,
\eeq
we find
\beq
 (\theta^n )_c =  -{(2n)! \over (2 i)^{2 n}} Y_n  \ .
\eeq
The first few cumulants are
\bea
(\th^2)_c = \oh Y_1 ~~~,~~~ (\th^4)_c = -{3\over 2} Y_2 ~~~,~~~ (\th^6)_c = {45 \over 4} Y_3  \ .
\label{first_cumulants}
\eea

The important point to notice in eq.\ \rf{rhopq} is that $\rho_{pq}(\th)$ will only be a Gaussian function of $\th$, and the higher cumulants will only vanish, if all the $Y_{n>1}=0$, which means that all the $X_{n>1}=0$.  As we will see in subsequent sections, it is not true that these quantities vanish, so the distribution must have non-Gaussian corrections.  Moreover, there is no reason to think that the $Y_{n>1}$ are negligible, for small quark masses, large $\b$, and sizeable chemical potential $\m$.   But how can this be consistent with the numerical results of \cite{Nakagawa:2011eu,Ejiri:2007ga}, which find that $\rho_{pq}(\th)$ seems to perfectly fit a Gaussian, even in the neighborhood of a phase transition, and moreover that $B_4=3$ within errors?

   The key to resolving this apparent inconsistency is the fact that all of the $X_n$, and all of the $Y_n$, are extensive quantities, proportional to the lattice volume $V$. 

\subsection{The apparent Gaussian shape of a non-Gaussian distribution}

    Observe that
\beq
\exp\left[-\sum_{n=1}^\infty  q^{2n} Y_n \right]  =  e^{-q^2 Y_1} \exp\left[-\sum_{n=2}^\infty  q^{2n} Y_n \right]  
\label{exp}
\eeq
is a Gaussian function of $q$ multiplied by non-Gaussian terms.  If all the $Y_{n>1}$ were zero, then $\rho_{pq}(\th)$ would also be a simple Gaussian function of $\th$.  The central region of $e^{-q^2 Y_1}$ is the region in which $q^2 Y_1$ is $O(1)$.  But in this region, where $q^2 \sim 1/Y_1$,
\beq
           q^{2n} Y_n \sim  {Y_n \over Y_1^n} \sim {1 \over V^{n-1}}  \ ,
\label{q2n}
\eeq
since all of the $Y_n \propto V$ are extensive.  This means that in the central region all of the $q^{2n} Y_n$ with $n>1$ are very small, on the order of powers of inverse volume, so inside that region we may expand the corresponding exponentials in a Taylor series
\bea
e^{-q^2 Y_1} \exp\left[-\sum_{n=2}^\infty  q^{2n} Y_n \right]  = e^{-q^2 Y_1}\bigg\{ 1 - Y_2 q^4 - Y_3 q^6 + 
(\oh Y_2^2 - Y_4) q^8 + ...\bigg\}  \ .
\label{taylor1}
\eea
What about outside the central region?  The higher order terms in the exponent of \rf{exp} are only of the same order of magnitude as $q^2 Y_1$ when $q^2$ becomes of $O(1)$ or larger, at which point the exponential is of order 
$e^{-cV}$. Thus the integrand in \rf{rhopq} is negligible outside the central region.  On the other hand, since the signs of the $Y_n$ are not necessarily all positive, one might wonder whether
\beq
          F(q^2) =  \sum_{n=1}^\infty  q^{2n} Y_n
\eeq
has minima with $F(q^2)<0$ away from $q^2=0$, in which case there would be an $e^{cV}$ \emph{enhancement}, and the integral would actually be dominated by such minima.  However, it is not hard to see that additional minima of this sort would lead to a breakdown of positivity in $\rho_{pq}(\th)$,\footnote{Just do a series expansion of the exponent around each of the hypothetical minima.  One obtains an oscillating function.} which, from the definition of this quantity, cannot occur. We can therefore ignore this possibility.    

   Inserting the Taylor expansion \rf{taylor1} into \rf{rhopq}, we integrate over $q$ and arrive at an expression for 
$\rho_{pq}(\th)$ which should be valid in the central region.  In order to display explicitly the volume dependence, we will write 
$Y_n = y_n V$, where the $y_n$ are all $O(1)$ in the volume.  Also, to avoid very long expressions, we will truncate the Taylor series in \rf{taylor1} at $O(q^6)$.  Carrying out the $q$-integration, we find for the angular distribution
\bea
\rho_{pq}(\th) &=& {1\over \sqrt{\pi y_1 V} }e^{-\th^2/(y_1 V)}\bigg( 1-\frac{3 y_2}{4 y_1^2}{1\over V}+\frac{3 \left(8 \th^2 y_2-5
   y_3\right)}{8 y_1^3}{1\over V^2}+\frac{45 \th^2 y_3-4 \th^4
   y_2}{4 y_1^4}{1\over V^3}
\non \\
& &   -\frac{15 \th^4 y_3}{2
   y_1^5}{1\over V^4}+\frac{\th^6 y_3}{
   y_1^6}{1\over V^5}+O(V^{-6})\bigg)  \ .
\label{rhopq1}
\eea  

   The pattern is clear: the phase angle distribution is a Gaussian in the central region, up to corrections which are expressed in a power series in $1/V$.\footnote{Of course it is understood that this is an asymptotic expansion, but the conclusion that the deviation from Gaussian is at most $O(1/V)$ can already be seen from \rf{q2n}.}   So at large volume it will be very difficult to distinguish $\rho_{pq}(\th)$ from a pure Gaussian.  But this fact does \emph{not} mean that the non-Gaussian corrections can be dropped.  In fact, it is these small corrections which are responsible for cumulants of 
4th-order and higher, which are not necessarily negligible; their magnitudes simply depend on the $y_n$.  To see this
explicitly, one can insert the expansion \rf{rhopq1} into the expression for the moments
\beq
\langle \th^n \rangle_{pq} = \int  d\th ~ \th^n \rho_{pq}(\th) \ ,
\eeq
and from the moments compute the cumulants.  The result for the first three cumulants is precisely the same as in \rf{first_cumulants}, which means that {\sl the higher-order cumulants are indeed coming from the inverse-volume corrections to the Gaussian form}, and contribute to cumulant expansion \rf{cumulant_expansion}.

\begin{figure}[t!]
\subfigure[~$V=2$]  
{   
 \label{V2}
 \includegraphics[scale=0.6]{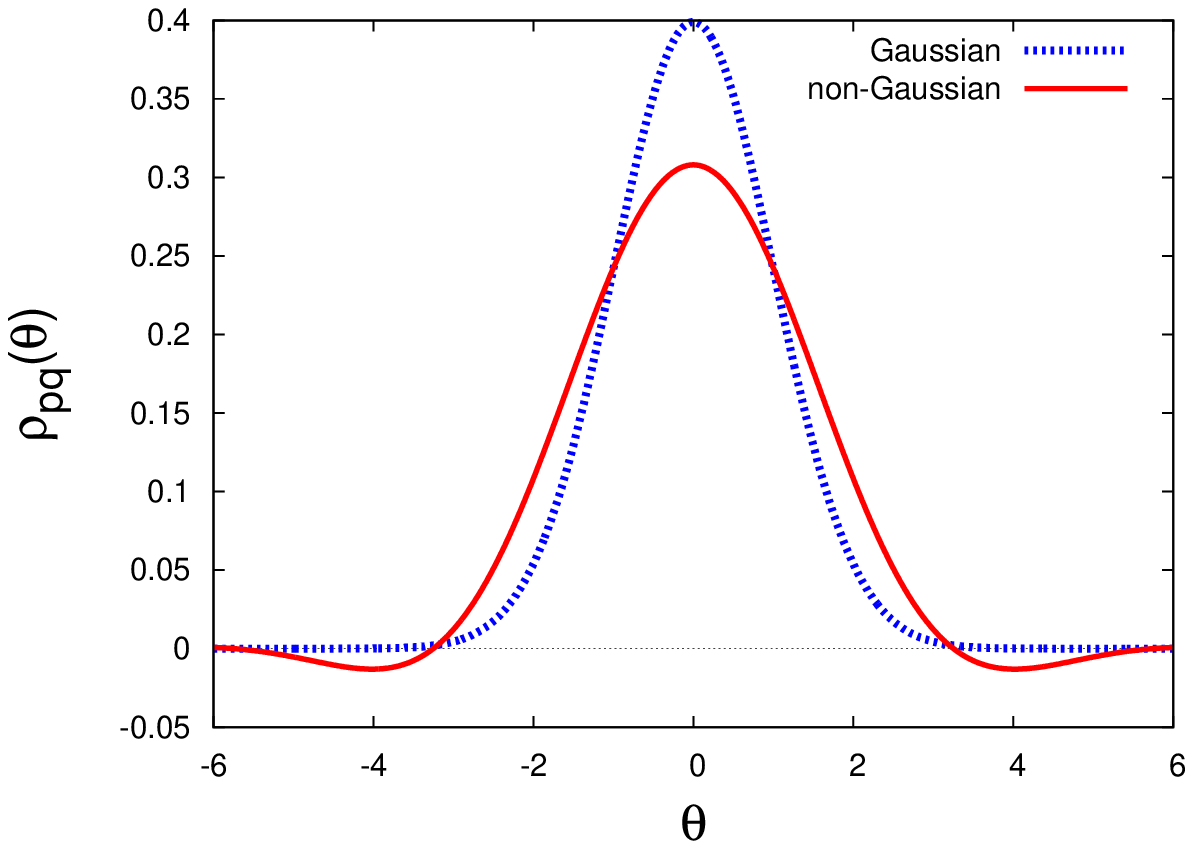}
}
\subfigure[~$V=10$]  
{   
 \label{V10}
 \includegraphics[scale=0.6]{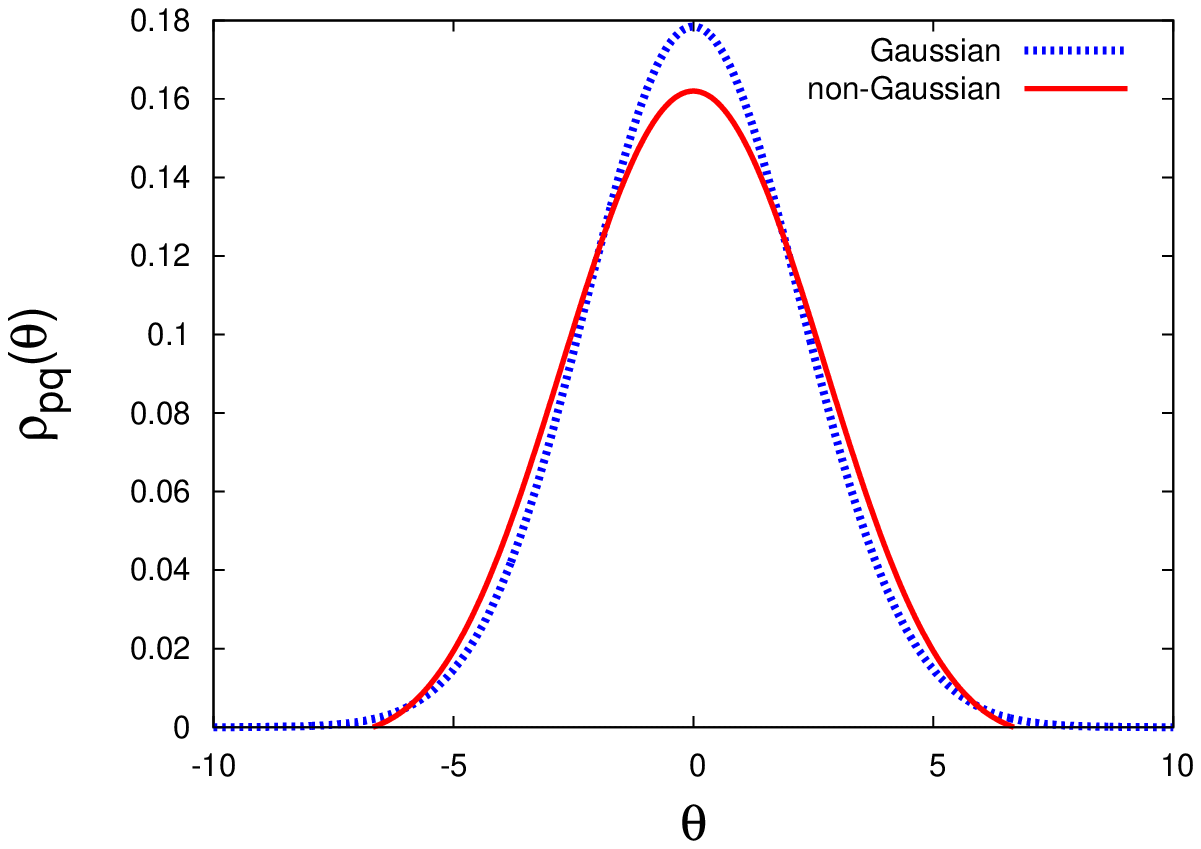}
}
\subfigure[~$V=50$]  
{   
 \label{V50}
 \includegraphics[scale=0.6]{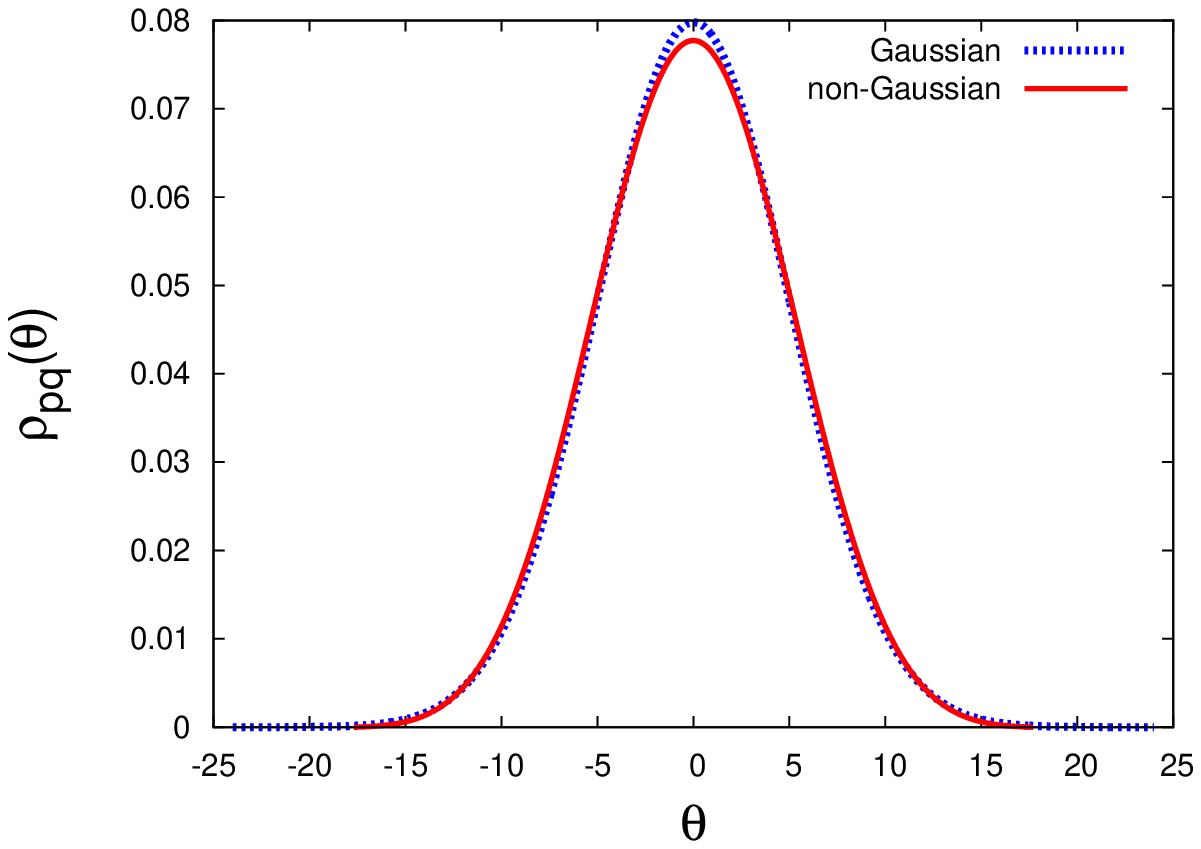}
}
\subfigure[~$V=250$]  
{   
 \label{V250}
 \includegraphics[scale=0.6]{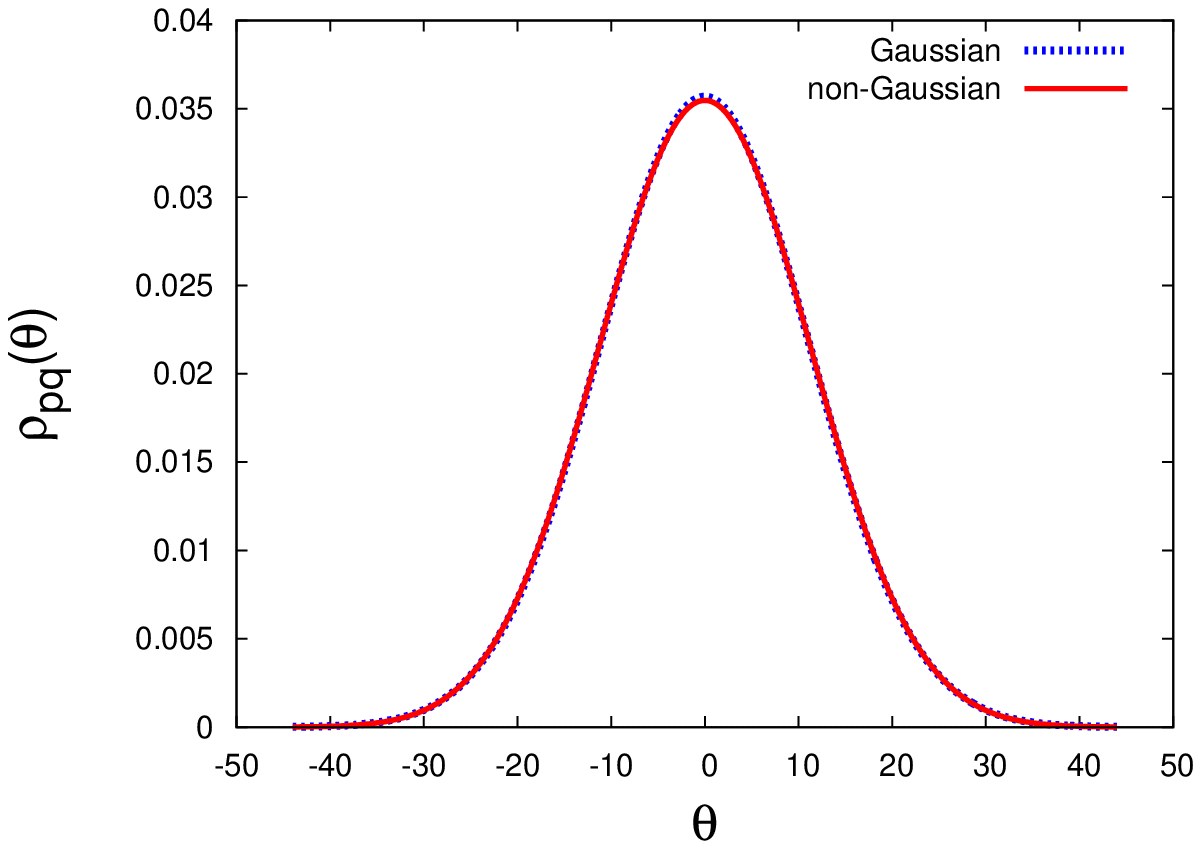}
}
\caption{The dashed blue curve is a Gaussian, obtained from \rf{rho} using Fourier components 
${\langle e^{2iq \th}\rangle_{pq} = \exp[-q^2 V]}$, while the red curve is a non-Gaussian distribution, with Fourier components
$\exp[-(q^2 + 2 q^4)V]$.  In the Gaussian case only the 2nd order cumulant non-zero, while in
the non-Gaussian case we have the ratio $(\th^4)_c/(\th^2)_c = -3$. However, as volume increases from $V=2$ to $V=250$, the red and blue curves become indistinguishable by eye.  Despite this apparent convergence, $(\th^4)_c=-3(\th^2)_c$ in the non-Gaussian distribution in all four cases, while
$(\th^4)_c=0$ for the Gaussian distribution.  The difference can only be attributed to an $O(1/V)$ deviation of the non-Gaussian distribution relative to the Gaussian distribution in the central region.}
\label{V}
\end{figure}

    As a simple illustration of these conclusions, let us take $y_1=1,~y_2=2,~y_{n>2}=0$, which means that the magnitude of the 4th-order cumulant is three times greater than that of the 2nd-order cumulant.  The resulting phase distribution, compared to the Gaussian distribution with $y_2$ set to zero, is shown for volumes $V=2,10,50,250$ in Fig.\ \ref{V}.  We see that at $V=250$ the Gaussian distribution and the actual distribution are almost indistinguishable by eye, despite the fact that 
$(\th^4)_c/(\th^2)_c = -3$ for all four volumes.\footnote{In this simple example, positivity is violated for the non-Gaussian curve. In a more realistic case, since $\r_{pq}(\th)$ is a probability density, the set of coefficients $\{y_n\}$ must satisfy the condition that $\r_{pq}(\th) \ge 0$.  But the point we are making here, i.e.\ that the Gaussian and non-Gaussian curves converge as 
$V\ra \infty$ while the cumulants do not, is very general, and does not depend on a specific choice of coefficients.}  

   Now we return to the fact that the Binder cumulant $B_4$ in (\ref{b4}) has been measured numerically, and the result appears
consistent with $B_4=3$, within error bars \cite{Ejiri:2007ga}.  But now we may ask how small those error bars have to be, and how accurately one must measure the 2nd and 4th-order moments, in order to rule out a significant 4-th order cumulant.
Since all the cumulants are proportional to one power of the volume, we have
\beq 
      (\th^4)_c  =   \langle \th^4 \rangle_{pq} - 3 \langle \th^2 \rangle_{pq}^2 =  O(V)
\eeq
On the other hand, both $\langle \th^4 \rangle_{pq}$ and $\langle \th^2 \rangle_{pq}^2$ are $O(V^2)$.  So this means that
\beq
              (\th^4)_c  =  (B_4 - 3) \langle \th^2 \rangle_{pq}^2 =  O(V)  \ ,
\eeq
in which case
\beq
                   B_4 - 3  =  O(1/V)  \ .
\eeq
Therefore, unless one has measured $B_4$ to an accuracy of $O(1/V)$ or better, one cannot rule out the possibility that
$O(1/V)$ deviations from $B_4=3$ result in significant 4th-order (and higher) cumulants, contributing to 
$\langle e^{iN_f \th} \rangle_{pq}$.

    Finally, it has been argued \cite{Ejiri:2007ga} that the Gaussian form of $\r_{pq}(\th)$ is a simple consequence of the Central
Limit Theorem.  The idea is that $\th$ is a sum of nearly independent contributions from regions of the lattice which are separated by a correlation length or more.  Suppose there are $N$ such contributions, proportional to the lattice volume, and each contribution is a stochastic variable with some probability distribution.  The Central Limit Theorem tells us that
the probability distribution for the sum of such contributions will \emph{approach} a Gaussian distribution as $N\ra \infty$.
But at any finite $N$ the probability distribution is not precisely a Gaussian, and in fact the Berry-Esseen Theorem places some bounds on the rate at which such a distribution approaches the Gaussian form.  These facts are completely consistent with the
points we are making here.  The distribution $\r_{pq}(\th)$ does approach a Gaussian as $V \ra \infty$, but in general there are $O(1/V)$ deviations from the Gaussian form which, as we have seen, can lead to higher-order cumulants even in the $V \ra \infty$ limit.

We now proceed to the detailed calculation of the $X_n$ in the hadron resonance gas, and in the strong coupling/hopping parameter expansion of lattice QCD.

\section{Free hadron gas}
\label{hrg}

Consider a gas of free mesons and baryons in the ground state. This includes mesons with spin $0$ and $1$ (spin degeneracy $g = 1$, $3$), and baryons with spin $\frac{1}{2}$ and $\frac{3}{2}$ (spin degeneracy $g = 2$, $4$). Each hadron has a contribution from the chemical potential given by
\EQ{
B \mu_B + 2 I_3 \mu_I \, ,
}
making use of the conventions where
\begin{equation*}
\begin{aligned}
& B \equiv \frac{1}{3} (N_q - N_{{\bar q}}) ~\text{is the baryon number},\\
& \mu_B = \mu_u + \mu_d + ... ~\text{is the baryon chemical potential},\\
& I_3 \equiv \frac{1}{2} \left[ \left( N_u - N_{{\bar u}} \right) - \left( N_d - N_{{\bar d}} \right) \right] ~ \text{is the third isospin component, and}\\
& \mu_I = \frac{1}{2} \left( \mu_u - \mu_d \right)~ \text{is the isospin chemical potential.}
\end{aligned}
\end{equation*}
For free mesons ($M$, ${\bar M}$) with mass $m_M$ and spin degeneracy $g = 2s+1$ the free energy $G_g^{M{\bar M}(\mu)}$ is given by \cite{Gross:1980br} (see also, for example \cite{Toublan:2004ks,Kapusta:2006pm,Myers:2009df})
\SP{
\ln Z_g^{M {\bar M}} &= g \frac{m_{M}^2 V_3 T}{\pi^2} \sum_{n=1}^{\infty} \frac{1}{n^2} {\rm K}_2(n m_{M}/T) \cosh \left[ 2 n I_3 \mu_I/T \right] \\
&= - \frac{1}{T} V_3 G_g^{M{\bar M}}(\mu_I) \, .
}
For free baryons ($B$, ${\bar B}$) with mass $m_B$ and spin degeneracy $g = 2s+1$ the free energy $G_g^{B{\bar B}(\mu)}$ is given by
\SP{
\ln Z_g^{B {\bar B}} &= - g \frac{m_B^2 V_3 T}{\pi^2} \sum_{n=1}^{\infty} \frac{(- 1)^n}{n^2} {\rm K}_2(n m_B/T) \cosh \left[ (\mu_B - 2 I_3 \mu_I )n /T \right] \\
&= - \frac{1}{T} V_3 G_g^{B{\bar B}}(\mu_B - 2 I_3 \mu_I) \, .
}

\subsection{Distribution of the complex phase}

To obtain the distribution of the complex phase it is necessary to calculate the expectation value on the r.h.s. of the equation for the moments in \rf{complex-phase},
\EQ{
\langle e^{2 i p \theta} \rangle_{QCD} = \frac{Z_{YM}}{Z} \bigg{\langle} \frac{\det^p (\Dsl+\gamma_0 \mu + m)}{\det^p(\Dsl-\gamma_0 \mu + m)} {\det}^{N_f} (\Dsl + \gamma_0 \mu + m) \bigg{\rangle} = \alpha_M(\mu) \alpha_B(\mu) \, ,
\label{phase-moms-hrg}
}
where $\alpha_M$ is the meson contribution and $\alpha_B$ is the baryon contribution, which can be separated assuming a non-interacting gas of hadrons. These are determined by collecting contributions from all possible mesons and baryons which can be formed from $2p+N_f$ quark flavors each with spin $\pm \frac{1}{2}$. The precise contributions are obtained by decomposing $SU(2(2p+N_f))$ to $SU(2p+N_f)_{flavor} \times SU(2)_{spin}$.

For mesons it is necessary to consider the decomposition of ${\bf n} \otimes {\bf \overline{n}}$ in $SU(n)$,
\SP{
{\bf n} \otimes {\bf \overline{n}} &= ({\bf n^2 - 1}) \oplus 1\\
&= {\bf \left( \left(\tfrac{n}{2}\right)^2 - 1\right)_{3} } \oplus {\bf \left( \left(\tfrac{n}{2}\right)^2 - 1\right)_{1} } \oplus {\bf 1}_{3} \oplus {\bf 1}_{1} \, ,
\label{meson-decomp}
}
where ${\bf R_g}$ is the decomposed representation with ${\bf R} \in SU(\frac{n}{2})$ and ${\bf g} \in SU(2)$. Thus the ${\bf \left( n^2 - 1\right) }$ and ${\bf 1}$ representations of flavor $SU(n)$ can have spin degeneracy $1$ or $3$ with equal likelihood.

For baryons consider the decomposition of ${\bf n} \otimes {\bf n} \otimes {\bf n}$ in $SU(n)$,
\SP{
{\bf n} \otimes {\bf n} \otimes {\bf n} = &\left( {\bf \frac{n(n-1)(n+1)}{3} } \right) \oplus \overline{\left({\bf \frac{n(n-1)(n-2)}{6}}\right)} \\
&\oplus \left( {\bf \frac{n(n-1)(n+1)}{3} } \right) \oplus \left( {\bf \frac{n(n+1)(n+2)}{6} } \right) \, .
\label{baryon-decomp}
}
The wavefunction of ground state baryons is spatially symmetric and color antisymmetric. Therefore it must be symmetric in flavor and spin combined and it is sufficient to consider the decomposition of the totally symmetric representation in (\ref{baryon-decomp}) to $SU(\frac{n}{2}) \times SU(2)$,
\EQ{
\left( {\bf \frac{n(n+1)(n+2)}{6} } \right) = \left( {\bf \frac{\frac{n}{2}(\frac{n}{2}-1)(\frac{n}{2}+1)}{3} } \right)_2 \oplus \left( {\bf \frac{\frac{n}{2}(\frac{n}{2}+1)(\frac{n}{2}+2)}{6} } \right)_4 \, .
\label{su2n-to-sun}
}
The r.h.s. shows that the $\left( {\bf \frac{n(n-1)(n+1)}{3} } \right)$ representation of flavor $SU(n)$ has a spin degeneracy $2$ and that the $\left( {\bf \frac{n(n+1)(n+2)}{6} } \right)$ has spin degeneracy $4$.

Since we are only including the contribution from mesons and baryons in the ground state it is sufficient to consider the direct products in flavor $SU(n)$ where each representation has the appropriate spin degeneracy determined from the decompositions above.

Using the results from the decomposition in (\ref{meson-decomp}), the meson contribution to \rf{phase-moms-hrg} is determined from the quark content of the weights (including multiplicities) of the $({\bf n^2 - 1})$ and the ${\bf 1}$ of flavor $SU(n)$, where each quark from the $\det^p$ in the denominator of (\ref{phase-moms-hrg}) comes with a factor of $-1$. The multiplicities are determined using Freudenthal's Recursion Formula
(see for example (XI.45) in \cite{Cahn:1985wk}). The result can be put in the form 
\beq
\left[ ((p+N_f)q_1-p q_2)((p+N_f){\bar q_1}-p{\bar q_2}) - N_f^2 q_1{\bar q_1} \right] \ ,
\eeq 
where $q_1$ is a quark contribution from a $\det$ in the numerator of (\ref{phase-moms-hrg}), and $q_2$ is a contribution from a $\det$ in the denominator. Each contribution of $q_1 {\bar q_1}$ or $q_2{\bar q_2}$ is associated with the free energy $\frac{1}{2}G^{M{\bar M}}(0)$, and each contribution of $q_1 {\bar q_2}$ or ${\bar q_1} q_2$ is associated with $\frac{1}{2}G^{M{\bar M}}(2 \mu)$, up to an overall multiplicative factor $k_M$. For an example of how this works see Appendix \ref{C}. The contribution to (\ref{phase-moms-hrg}) from the ground state mesons is then given by
\EQ{
\alpha_M(\mu) = \exp \left[ - p(p+N_f) \left[ G^{M{\bar M}}(2 \mu) - G^{M{\bar M}}(0) \right] \right] \, ,
\label{alphaM}
}
where from the meson decomposition in (\ref{meson-decomp}), the free energy is split evenly between $q=1$ and $g=3$ contributions, so
\EQ{
G^{M{\bar M}}(\mu) = \frac{1}{2} k_M \left[ G_1^{M{\bar M}}(\mu) + G_3^{M{\bar M}}(\mu) \right] \, .
}

Using the results from the decomposition in (\ref{su2n-to-sun}), the baryon contribution to (\ref{phase-moms-hrg}) is similarly determined from the quark content of the weights (including multiplicities) of the flavor $SU(n)$ representations $\left( {\bf \frac{n(n-1)(n+1)}{3} } \right)$ which has $g = 2$, $\left( {\bf \frac{n(n+1)(n+2)}{6} } \right)$ with $g = 4$, and $\left( {\bf \frac{n(n-1)(n-2)}{6} } \right)$ with $g = 4$ (required to obtain the identity from (\ref{phase-moms-hrg}) when $\mu = 0$). Adding separately the $g = 2$ and $g = 4$ baryon contributions the result can be obtained from $-\left[ ((p+N_f)q_1-p q_2)^3 - N_f^3 q_1^3 \right]$, where each contribution from $q_1^3$ or $q_2^3$ is associated with $G^{B{\bar B}}(3\mu)$ and each contribution from $q_1^2 q_2$ or $q_1 q_2^2$ is associated with $G^{B{\bar B}}(\mu)$, up to an overall multiplicative factor $k_B$. See Appendix \ref{C} for details. The contribution to (\ref{phase-moms-hrg}) from the ground state baryons is then
\EQ{
\alpha_B(\mu) = \exp \left[ - p(p+N_f) \left[ G^{B{\bar B}}(3 \mu) - G^{B{\bar B}}(\mu) \right] \right] \, ,
\label{alphaB}
}
where from the baryon decomposition in (\ref{baryon-decomp}), there are two contributions with $g = 2$ for each with $g = 4$, so
\EQ{
G^{B{\bar B}}(\mu) = k_B \left[ 2 G_2^{B{\bar B}}(\mu) + G_4^{B{\bar B}}(\mu) \right] \, .
}
The final form of (\ref{phase-moms-hrg}) is then
\SP{
\langle e^{2 i p \theta} \rangle_{QCD} &= \frac{Z_{YM}}{Z} \bigg{\langle} \frac{\det^p (\Dsl+\gamma_0 \mu + m)}{\det^p(\Dsl-\gamma_0 \mu + m)} {\det}^{N_f} (\Dsl + \gamma_0 \mu + m) \bigg{\rangle}\\
&= \alpha_M(\mu) \alpha_B(\mu)\\
&= e^{-p(p+N_f) X_1} \, ,
\label{complex-phase-moms-result}
}
where
\EQ{
X_1 = G^{M{\bar M}}(2 \mu) - G^{M{\bar M}}(0) + G^{B{\bar B}}(3 \mu) - G^{B{\bar B}}(\mu) \, .
\label{X-def}
}
In this case the $X_i$ with $i > 1$ are zero in the more general form of $\langle e^{2 i p \theta} \rangle_{QCD}$ in (\ref{form}), thus the integral over $p$ in (\ref{rho}) can be performed leading to the Gaussian form for the distribution 
$\langle \delta(\theta - \theta') \rangle_{QCD}$. We note that the distribution could not have the more general form due to an insufficient number of quark lines in the one-loop integral contributing to the free energy.

We note that the distribution of the baryon number, which leads to a description of the noise of the sign problem as a total derivative in \cite{Greensite:2013vza},  follows in a similar manner, and is presented in Appendix $\ref{D}$.

\section{Strong coupling expansion}
\label{sc}

The partition function for Yang-Mills theory from the lattice strong coupling expansion can be obtained by integrating out the spatial link variables.  This results in an expression for the effective action in terms of a character expansion of the Polyakov line and its adjoint (see for example, \cite{Drouffe:1983fv,Green:1983sd,Montvay:1994cy})
\SP{
Z_{YM} &= \int_{SU(N_c)} \prod_z {\rm d}W_z \prod_{\langle x y \rangle} \left[ 1 + \sum_{R} \lambda_R \left[ \chi_R(W_x)\chi_R(W_{y}^{\dagger}) + \chi_{R}(W_{x}^{\dagger}) \chi_R(W_y) \right] \right] \, ,
}
where $\langle x y \rangle$ indicates that the product extends over all nearest neighbor sites $x$ and $y$, and the coupling dependence is contained within the representation dependent prefactor $\lambda_{R} = \lambda_{R}(\beta_{lat})$ with $\beta_{lat} = \frac{2 N_c}{g^2}$. $\chi_R(W_{\bf x})$ is the character in the representation $R$ of the Polyakov line $W_{\bf x}$ at site ${\bf x}$, which corresponds to the ordered product of links in the temporal direction,
\SP{
W_{{\bf x}} &= \prod_{\tau = 1}^{N_{\tau}} U_0({\bf x},\tau) \, .
}
The first nontrivial contribution to the partition function, as the coupling $g^2 \rightarrow \infty$, is obtained by truncating the sum over $R$ after the fundamental representation,
\SP{
Z_{YM} &\xrightarrow[g^2 \rightarrow \infty]{} \int_{SU(N_c)} \prod_z {\rm d}W_z \prod_{\langle x y \rangle} \left[ 1 + \lambda_F \left[ \tr(W_x) \tr(W_{y}^{\dagger}) + \tr (W_{x}^{\dagger}) \tr (W_y) \right] \right] \, ,
\label{Z_YM_lambda}
}
where \cite{Drouffe:1983fv,Montvay:1994cy}
\EQ{
\lambda_F = \left[ u \right]^{N_{\tau}} \left[ 1 + {\cal O}\left( u \right) \right] \, ,
}
with $u \equiv \frac{1}{3} \left[ \frac{x}{2} + \frac{x^2}{8} - ... \right]$ for $N_c = 3$ and $x \equiv \frac{2}{g^2}$. In the subsequent subsections our calculations follow the procedure and use the techniques developed in \cite{Fromm:2011qi,Langelage:2008dj,Langelage:2009jb,Langelage:2010yn,Langelage:2010yr,Fromm:2012eb}.

\subsection{Hopping expansion}

To obtain the distribution of the complex phase $\rho(\theta')$, it is necessary to calculate the moments $\langle e^{2 i p \theta} \rangle_{QCD}$ from the expectation value in \rf{complex-phase}, which can be evaluated via a hopping expansion. We will compute the $\mu$-dependent part of $Q \equiv \frac{Z}{Z_{YM}} \langle e^{2 i p \theta} \rangle_{QCD}$, which takes the form
\SP{
Q &\equiv \bigg{\langle} \frac{\det^p (\Dsl + \gamma_0 \mu + m)}{\det^p (\Dsl - \gamma_0 \mu + m)} \det^{N_f} (\Dsl + \gamma_0 \mu + m) \bigg{\rangle} \\
&= 1 + q_1 h + q_2 h^2 + q_3 h^3 + ... \, ,
\label{Q-expansion}
}
where $h$ is an expansion parameter, dependent on temperature and on the specific form of the lattice action, which goes to zero as quark mass goes to infinity.  It will be specified below. Since $Q$ is a partition function it is natural that the final result has the form
\EQ{
Q = \exp \left[ F(\mu) V \right] \, ,
\label{lin-in-vol}
}
where $V$ is the volume of a three-dimensional time-slice of the lattice, and $F(\mu)$ can be expressed as a series in $h$. The expansion (\ref{Q-expansion}) contains various powers of $V$, which can be understood as a Taylor series expansion of (\ref{lin-in-vol}).   In each coefficient $q_i$ we look for the piece which is linear in $V$, and these terms together give us $F(\mu)$ as a power series in $h$.

To illustrate the general form of $Q$ at large quark mass, it is sufficient to consider the Dirac operator of the naive fermion action with a chemical potential \cite{Hasenfratz:1983ba}
\SP{
(\Dsl + \gamma_0 \mu + m)_{x y} &= m \delta_{x y} \mathbbm{1} + \frac{1}{2} \sum_{k = 1}^{3} \gamma_k \left[ U_k(x) \delta_{x+{\hat k},y}  - U_k^{\dagger}(y) \delta_{x - {\hat k},y} \right] \\
&\hspace{17mm}+ \frac{1}{2} \gamma_4 \left[ e^{\mu} U_k(x) \delta_{x+{\hat 4},y} - e^{-\mu} U_k^{\dagger}(y) \delta_{x-{\hat 4},y} \right] \\
&= m \delta_{x y} \mathbbm{1} + R(\mu)_{x y} \, .
}
Since $m$ is assumed to be very large in lattice units, we may write
\SP{
\Tr \log (\Dsl + \gamma_0 \mu + m) &= \Tr \log \left[ m \mathbbm{1} + R(\mu) \right] \\
&= \Tr \log \left[ \mathbbm{1} + \frac{1}{m} R(\mu) \right] + \Tr \log \left[ m \mathbbm{1} \right] \\
&= \sum_{n=1}^{\infty} \left( \frac{1}{m} \right)^{n} (-1)^{n+1} (-1)^{l+1} \frac{1}{n} \Tr R^n(\mu) + \Tr \log \left[ m \mathbbm{1} \right] \, .
}
The last term is an irrelevant, $\mu$-independent constant, which we will drop from here on. The trace is understood to sum over both Dirac indices and position, and $\Tr R^n$ is given by closed Wilson loops times some numerical factors. We note that there is an additional factor of $(-1)^{l+1}$ to incorporate the effect of anti-periodic boundary conditions on fermions, where $l$ is the number of temporal windings of $R^n$.

Things simplify because we are only interested in the $\mu$-dependent terms. This tells us that we need only consider loops with non-zero winding number around the compact time direction. Note that
\SP{
&\frac{1}{N_t} \Tr R^{N_t}(\mu) \\
&= \frac{1}{N_t} \sum_x \sum_{t=1}^{N_t} e^{\mu N_t} \Tr \left[ U_0({\bf x},t) U_0 ({\bf x},t+1) ... U_0({\bf x},t+N_t-1) \right] \left( \frac{1}{2} \right)^{N_t} \Tr \left[ \gamma_4^{N_t} \right] \\
&\hspace{6mm}+ \frac{1}{N_t} \sum_{{\bf x}} \sum_{t=1}^{N_t} e^{-\mu N_t} \Tr \left[ U_0({\bf x},t) U_0({\bf x}, t+1) ... U_0({\bf x}, t+N_t - 1) \right]^{\dagger} \left( \frac{1}{2} \right)^{\dagger} \Tr \left( - \gamma_4 \right)^{N_t}
}
where $N_t$ is the number of temporal lattice slices. Since the trace is cyclic-symmetric, we can just pick $t = 1$ in the sum over $t$ and drop the factor of $\frac{1}{N_t}$. Also, since the trace over an odd power of gamma matrices is zero, we only consider $N_t$ even (this requirement is special for the naive action). So we have
\EQ{
\frac{1}{N_t} \Tr R^{N_t}(\mu) = \left( \frac{1}{2} \right)^{N_t} 4 \sum_{{\bf x}} \left( e^{\mu N_t} P_{{\bf x}} + e^{- \mu N_t} P_{{\bf x}}^{\dagger} \right)
}
where $P_{\bf x}$ is the trace of the Polyakov line holonomy $W_{\bf x}$. Therefore, in the heavy quark limit the $\mu$-dependent part of the fermion contribution to the action has reduced to the general form
\SP{
\log \det (\Dsl + \gamma_0 \mu + m) = ~&a_1 h (e^{\mu/T} L_1 + e^{-\mu/T} L_1^*) + a_2 h^2 (e^{2 \mu/T} L_2 + e^{-2 \mu/T} L_2^*) \\
&+ a_3 h^3 (e^{3 \mu/T} L_3 + e^{-3 \mu/T} L_3^*) + ... \, ,
\label{ferm-action}
}
with
\SP{
L_n &\equiv \sum_{{\bf x}} \tr W_{{\bf x}}^n = \sum_{{\bf x}} P_{\bf x}^n \, , \\
L_n^* &\equiv \sum_{{\bf x}} \tr (W_{{\bf x}}^{\dagger})^{n} = \sum_{{\bf x}} P_{\bf x}^{\dagger n} \, .
}
The ${\cal O}(1)$ constants $a_n$ and the hopping parameter $h$ depend on the lattice quark action. For the naive fermion action $a_n = {4\over n}(-1)^n$.  Note that the general form in (\ref{ferm-action}) agrees with the fermion contribution to the effective action obtained from the hopping expansion for Wilson fermions where $a_n = \frac{2}{n} (-1)^n$ \cite{Fromm:2011qi}. The expansion parameter $h$ is $\left(\frac{1}{2m}\right)^{N_t}$ in the case of the naive action, while in the case of the Wilson action this parameter is $h \equiv (2 \kappa_f)^{N_t}$, with $\kappa_f = \frac{1}{D+1+m a}$, $D = 3$ spatial dimensions, and lattice spacing $a$. It is now possible to write $Q$ in (\ref{Q-expansion}) as the VEV of an exponential of terms of the form shown in (\ref{ferm-action}).  Expanding the exponential in a powers series in $h$, and taking the expectation value, will determine the coefficients $q_i$ shown in (\ref{Q-expansion}).  In \cite{Langelage:2010yn} the authors found that taking $a_1 = -2$, $a_2 = 1$, $a_3 = -\frac{2}{3}$ and truncating the rest allows for a direct comparison with the hadron resonance gas model.  Since the calculation in \cite{Langelage:2010yn} is ${\cal O}(h^3)$, our results are consistent.  The differences we find between the strong-coupling/hopping parameter expansion and the hadron resonance gas model only appear at ${\cal O}(h^4)$.

\subsection{${\cal O}(h^2)$}

Since all expectation values are with respect to the Yang-Mills vacuum, $\langle L_n^j \rangle = \langle (L_n^*)^j \rangle$ is used and we expand $Q$ up to ${\cal O}(h^2)$ to obtain
\EQ{
q_1 = 2 a_1 N_f \cosh(\mu/T) \langle L_1 \rangle \, ,
\label{q1h2}
}
\SP{
q_2 = &2 a_1^2 p^2 \left[ \cosh(2\mu/T) - 1 \right] \left[ \langle L_1^2 \rangle - \langle L_1 L_1^* \rangle \right] \\
&+ 2 a_1^2 p N_f \left[ \cosh(2\mu/T) - 1 \right] \left[ \langle L_1^2 \rangle - \langle L_1 L_1^* \rangle \right] \\
&+ 2 a_2 N_f \cosh(2 \mu/T) \langle L_2 \rangle + a_1^2 N_f^2 \left[ \cosh(2\mu/T) \langle L_1^2 \rangle + \langle L_1 L_1^* \rangle \right] \, .
\label{q2h2}
}
Therefore the leading contribution to $X_1$ in $\langle e^{2 i p \theta} \rangle_{QCD}$ in the format of (\ref{form}) is
\EQ{
X_1 = 2 a_1^2 h^2 [\cosh(2\mu/T) - 1] \left[\langle L_1 L_1^* \rangle - \langle L_1^2 \rangle \right] + {\cal O}(h^3) \, ,
\label{X1h2}
}
where terms without any $p$-dependence have been cancelled off by the partition function in the denominator. Note that in the confined phase $\langle L_1 \rangle = \langle L_1^2 \rangle = 0$ and $\langle L_1 L_1^* \rangle = N_s$, where $N_s$ is the number of spatial lattice sites, so one can solve (\ref{X1h2}) to obtain $X_1 = 2 a_1^2 h^2 N_s [\cosh(2\mu/T) - 1] + {\cal O}(h^3)$. Therefore, up to ${\cal O}(h^2)$, the simpler structure $Q \frac{Z_{YM}}{Z} = \exp[-p(p+N_f)X_1]$ is maintained. We will now show that this is no longer the case at ${\cal O}(h^4)$.

\subsection{${\cal O}(h^4)$ corrections}

Considering the ${\cal O}(h^4)$ contribution to $Q$ and keeping the ${\cal O}(p^3)$ and ${\cal O}(p^4)$ contributions we obtain
\EQ{
q_4 = \frac{4}{3}a_1^4 (p^4 + 2 p^3 N_f) \sinh^4(\mu/T) \left[ \langle L_1^4 \rangle - 4 \langle L_1^3 L_1^* \rangle + 3 \langle L_1^2 L_1^{* 2} \rangle \right] + {\cal O}(p^2) \, .
\label{q4}
}
From here on we focus on obtaining results in the confined phase. We work to leading order in the limit that the coupling $g^2 \rightarrow \infty$ using the effective gauge action in (\ref{Z_YM_lambda}),
\EQ{
e^{-S_{eff}} = \prod_{\langle x y \rangle} \left[ 1 + \lambda_1 \left[ \tr(W_x) \tr(W_{y}^{\dagger}) + \tr (W_{x}^{\dagger}) \tr (W_y) \right] \right] \, ,
}
where $\lambda_1 = \left(\frac{1}{g^2 N_c}\right)^{N_{\tau}}$. In this case the only expectation values which survive are those which involve $SU(3)$ color singlet contributions (see, e.g., chapter $8$ in \cite{Cvitanovic:2008zz}). For general $N_c$ the surviving expectation values have the form
\EQ{
\int_{SU(N_c)} {\rm d}W~ (\tr W \tr W^{\dagger})^l (\tr W)^{N_c m} (\tr W^{\dagger})^{N_c n} \ne 0 \, ,
}
where $l$, $m$, $n = 0, 1, 2, ...$. This allows us to obtain the following expectation values for general $N_c$
\EQ{
\langle P_x P_x^* \rangle = 1 \, ,
\label{ppstar}
}
\EQ{
\langle P_x^2 P_x^{* 2} \rangle = 2 \, ,
\label{ppstar2}
}
and for $N_c = 3$
\EQ{
\langle P_x^3 \rangle = 1 \, ,
}
where the details are worked out in Appendix \ref{E}. Using these results considering $N_c = 3$ we obtain
\begin{align}
\langle L_1^4 \rangle = &~0 \, , \\
\langle L_1^3 L_1^* \rangle = &~0 \, , \\
\label{h4_disc1}
\langle L_1^2 L_1^{* 2} \rangle = &~2 \sum_{x \ne y} \langle P_x P_y P_x^* P_y^* \rangle + \sum_x \langle P_x^2 P_x^{* 2} \rangle + { \sum_{x \ne y} \langle P_x P_x P_y^* P_y^* \rangle} + 4 \sum_{x \ne y} \langle P_x P_x P_x^* P_y^* \rangle \nonumber\\
&~+ 4 \sum_{x \ne y \ne z} \langle P_x P_y P_x^* P_z^* \rangle + {\cal O}(\lambda_1^2) \\
= &~2 \sum_{x \ne y} + 2 \sum_x + { 18} \lambda_1 \sum_{\langle x y \rangle} + 8 \lambda_1 \sum_{x \ne \langle y z \rangle}\, + {\cal O}(\lambda_1^2) \nonumber\\
= &~2 N_s^2 + { 6} N_s \lambda_1 + 24 N_s^2 \lambda_1 + {\cal O}(\lambda_1^2) \, , \nonumber
\end{align}
recalling that $N_s$ is the number of spatial lattice sites. Using these results in (\ref{q4}) leads to a contribution to $X_2$ of the form $X_2 = - \frac{4}{3} a_1^4 h^4 \sinh^4(\mu/T) \left[ 18 \lambda_1 N_s \right] + {\cal O}(h^5)$. Note that if $\lambda_1 = 0$ then there are no contributions to $X_2$. However, at ${\cal O}(\lambda_1)$, there is a contribution at ${\cal O}(V)$ (proportional to $N_s$) which arises in (\ref{h4_disc1}) because $\sum_{x \ne y} \langle P_x P_x P_y^* P_y^* \rangle \ne 0$ for $SU(3)$.  Notice that $\sum_{x \ne y} \langle P_x P_x P_y^* P_y^* \rangle$ is zero in the limit $N_c \rightarrow \infty$ so in that case there is no ${\cal O}(V)$ term and the simpler $Q \frac{Z_{YM}}{Z} = \exp[-p(p+N_f)X_1]$ structure is maintained. In the next section we show that at ${\cal O}(h^6)$ the distribution for $N_c = 3$ also acquires corrections to the Gaussian form at ${\cal O}(\lambda_1^0)$.

\subsection{${\cal O}(h^6)$ corrections}

It is straightforward, though a bit tedious, to calculate the contributions to $Q$ at ${\cal O}(h^6 p^6)$ and ${\cal O}(h^6 p^5)$. The result is,
\EQ{
q_6 = \frac{8}{45} a_1^6 \sinh^6(\mu/T) \left[ \langle L_1^6 \rangle - 6 \langle L_1^5 L_1^* \rangle + 15 \langle L_1^4 L_1^{* 2} \rangle - 10 \langle L_1^3 L_1^{* 3} \rangle \right] p^5 \left( p + 3 N_f \right) + {\cal O}(p^4) \, .
\label{q6}
}
To solve for $q_6$ the following expectation values are required in addition to (\ref{ppstar}) and (\ref{ppstar2}): for $SU(3)$
\SP{
\langle P_x^6 \rangle &= 5 \, , \\
\langle P_x^4 P_x^* \rangle &= 3 \, ,
}
for $SU(N_c)$
\EQ{
\langle P_x^3 P_x^{* 3} \rangle &= 6 \, . \\
}
For $SU(3)$ the expectation values in (\ref{q6}) are then
\SP{
\langle L_1^6 \rangle &= -5 N_s + 10 N_s^2 - 270 N_s \lambda_1 + 360 N_s^2 \lambda_1 \, , \\
\langle L_1^5 L_1^* \rangle &= 0 \, , \\
\langle L_1^4 L_1^{* 2} \rangle &= 0 \, , \\
\langle L_1^3 L_1^{* 3} \rangle &= -N_s +N_s^2 + 6 N_s^3 - 72 N_s \lambda_1 + 90 N_s^2 \lambda_1 + 108 N_s^3 \lambda_1 \, ,
}
up to ${\cal O}(\lambda_1^2)$ corrections, resulting in
\SP{
&\langle L_1^6 \rangle - 6 \langle L_1^5 L_1^* \rangle + 15 \langle L_1^4 L_1^{* 2} \rangle - 10 \langle L_1^3 L_1^{* 3} \rangle \\
&\hspace{2cm}= { 5} N_s - 60 N_s^3 + { 450} N_s \lambda_1 - 540 N_s^2 \lambda_1 - 1080 N_s^3 \lambda_1 + {\cal O}(\lambda_1^2)\, .
}
This indicates that there are $O(V)$ contributions to $Q$ at ${\cal O}(h^6 p^6)$ and ${\cal O}(h^6 p^5)$, at $O(\lambda_1^0)$ and ${\cal O}(\lambda_1)$, at least for $SU(3)$.

For $SU(N_c)$ in the limit $N_c \rightarrow \infty$, one obtains, up to $O(\lambda_1^2)$ corrections,
\SP{
\langle L_1^6 \rangle &= 0 \, , \\
\langle L_1^5 L_1^* \rangle &= 0 \, , \\
\langle L_1^4 L_1^{* 2} \rangle &= 0 \, , \\
\langle L_1^3 L_1^{* 3} \rangle &= 6 N_s^3 + 108 N_s^3 \lambda_1 \, ,
}
such that there are no contributions at $O(V)$ or $O(V^2)$ at $O(\lambda_1^0)$ or ${\cal O}(\lambda_1)$, in the limit $N_c \rightarrow \infty$.

\subsection{Summary of strong coupling results}

Including the results from the previous sections, our calculations provide evidence which supports a structure of the form $Q \frac{Z_{YM}}{Z} = \exp \left[ -p(p+N_f) X_1 \right]$, for $N_c \rightarrow \infty$. For $SU(3)$ there are corrections 
to this form starting at ${\cal O}(h^4)$ for $O(\lambda_1)$, and at ${\cal O}(h^6)$ for $O(\lambda_1^0)$. To summarize, in the confined phase we have found
\SP{
X_1 &= 2 a_1^2 h^2 \left[ \cosh(2\mu/T) - 1\right] \sum_{{\bf x}} \left[ \langle \tr W_{{\bf x}} \tr W_{{\bf x}}^{\dagger} \rangle - \langle (\tr W_{{\bf x}})^2 \rangle \right] + {\cal O}(h^3) \\
&= 4 a_1^2 h^2 N_s \sinh^2 (\mu/T) + {\cal O}(h^3) \, ,
}
which is the contribution at the Gaussian level,
\EQ{
X_2 = - \frac{4}{3} a_1^4 h^4 \sinh^4(\mu/T) \left[ 18 \lambda_1 N_s \right] + {\cal O}(h^5)\, ,
\label{x2}
}
which is a correction to the Gaussian distribution at ${\cal O}\left(\lambda_1 \right)$, and
\EQ{
X_3 = - \frac{8}{45} a_1^6 h^6 \sinh^6(\mu/T) \left[ (5 + 450 \lambda_1) N_s \right] + {\cal O}(h^7)\, ,
\label{x3}
}
which contains a correction at ${\cal O}(\lambda_1^0)$ and ${\cal O}(\lambda_1)$. We note that the corrections $X_2$, and $X_3$, could not have appeared at a lower order in $h$ because the number of winding loops is insufficient. This would also be true for the deconfined phase, though the $X_i$ take different values.

\section{Taylor expansion}
\label{taylor}

Our results from the strong coupling expansion for $N_c = 3$ suggest that non-Gaussian terms arise in QCD in the limit of strong coupling, and large quark mass, at ${\cal O}(h^4)$ in the hopping expansion.

In what follows we also show that by performing a Taylor expansion around $\mu = 0$, non-Gaussian terms necessarily become possible unless some special relationships exist between certain expectation values of derivatives of fermion determinants, as found in \cite{Ejiri:2009hq}. We note that from the strong coupling expansion the correction $X_2$ in (\ref{x2}) appears at ${\cal O}(\mu/T)^4$, and $X_3$ in (\ref{x3}) appears at ${\cal O}(\mu/T)^6$.

Let $M(\mu) \equiv \det [\Dsl + \gamma_0 \mu + m]$. Expansion of $Q \equiv \langle \frac{M(\mu)^{p+N_f}}{M(-\mu)^p} \rangle$ in a Taylor series around $\frac{\mu}{T} = 0$ leads to
\SP{
\bigg{\langle} \frac{M(\mu)^{p+N_f}}{M(-\mu)^p} \bigg{\rangle} \sim ~&\langle M(0)^{N_f} \rangle + \frac{1}{2!} \left( \frac{\mu}{T} \right)^2 \bigg{\langle} \frac{\partial^2 D(\mu)}{\partial (\mu/T)^2}\bigg{|}_{\mu=0} \bigg{\rangle} + \frac{1}{4!} \left( \frac{\mu}{T} \right)^4 \bigg{\langle} \frac{\partial^4 D(\mu)}{\partial (\mu/T)^4}\bigg{|}_{\mu=0} \bigg{\rangle} \\
&+ \frac{1}{6!} \left( \frac{\mu}{T} \right)^6 \bigg{\langle} \frac{\partial^6 D(\mu)}{\partial (\mu/T)^6} \bigg{|}_{\mu=0} \bigg{\rangle} + \frac{1}{8!} \left( \frac{\mu}{T} \right)^8 \bigg{\langle} \frac{\partial^8 D(\mu)}{\partial (\mu/T)^8} \bigg{|}_{\mu=0} \bigg{\rangle} + {\cal O}\left( \frac{\mu}{T} \right)^{10}
}
where the contributions from taking an odd number of derivatives with respect to $\frac{\mu}{T}$ are zero because the partition function is even in $\mu$. Defining $D(\mu) = \frac{M(\mu)^{p+N_f}}{M(-\mu)^{p}}$ the derivatives are
\EQ{
\frac{\partial^2 D(\mu)}{\partial (\mu/T)^2} \bigg{|}_{\mu=0} = 4 p(p+N_f) M(0)^{N_f - 2} M'(0)^2 + {\cal O}(p^0)
\label{d2}
}
\SP{
&\frac{\partial^4 D(\mu)}{\partial (\mu/T)^4} \bigg{|}_{\mu=0} = 16 p^2 (p+N_f)^2 M(0)^{N_f - 4} M'(0)^4 + 8 p (p+N_f) M(0)^{N_f - 4} M'(0)\\
&\hspace{10mm}\times \left[ (4 - 3 N_f + N_f^2) M'(0)^3 + 3 (N_f - 2) M(0) M'(0) M''(0) + 2 M(0) M'''(0) \right] \\
&\hspace{10mm}+ {\cal O}(p^0)
\label{d4}
}
\SP{
&\frac{\partial^6 D(\mu)}{\partial (\mu/T)^6} \bigg{|}_{\mu=0} = 64 p^3 (p+N_f)^3 M(0)^{N_f - 6} M'(0)^6 + 16 p^2 (p+N_f)^2 M(0)^{N_f - 6} M'(0)^3 \\
&\hspace{10mm}\times \left[ (40 - 15 N_f + 3 N_f^2) M'(0)^3 + 15(N_f - 4) M(0)M'(0)M''(0) + 20 M(0)^2 M'''(0) \right] \\
&\hspace{10mm}+ p (p+N_f) X_{6,1} + {\cal O}(p^0)
\label{d6}
}
\SP{
&\frac{\partial^8 D(\mu)}{\partial (\mu/T)^8} \bigg{|}_{\mu=0} = 256 p^4 (p+N_f)^4 M(0)^{N_f - 8} M'(0)^8 + 256 p^3 (p+N_f)^3 M(0)^{N_f - 8} M'(0)^5 \\
&\hspace{10mm}\times \left[ (28 - 7 N_f + N_f^2) M'(0)^3 + 7 (N_f - 6) M(0) M'(0) M''(0) + 14 M(0)^2 M'''(0) \right] \\
&\hspace{10mm}+ p^2 (p+N_f)^2 X_{8,2} + p(p+N_f) X_{8,1} + {\cal O}(p^0)
\label{d8}
}
The expressions labeled by $X_{n,k}$ are lengthy functions of $M(0)$ and derivatives of $M(\mu)$ evaluated at $\mu = 0$. They have no $p$ or $\mu$-dependence.

To determine how the series exponentiates consider the Taylor expansion of $\log \langle e^{2 i p \theta} \rangle = \log \frac{Q}{Z}$,
\SP{
\log \langle e^{2 i p \theta} \rangle_{QCD} = ~&\log \bigg[ \frac{Z_{YM}}{Z} \bigg{\langle} \frac{M(\mu)^{p+N_f}}{M(-\mu)^p} \bigg{\rangle} \bigg] \\
\sim ~&\frac{1}{2!} \left( \frac{\mu}{T} \right)^2 \left[ \frac{\langle D^{(2)}(0) \rangle}{\langle M(0)^{N_f} \rangle} \right] \\
&+ \frac{1}{4!} \left( \frac{\mu}{T} \right)^4 \left[ \frac{\langle D^{(4)}(0) \rangle}{\langle M(0)^{N_f} \rangle} - 3 \frac{\langle D^{(2)}(0) \rangle^2}{\langle M(0)^{N_f} \rangle^{2}} \right] \\
&+ \frac{1}{6!} \left( \frac{\mu}{T} \right)^6 \bigg{[} \frac{\langle D^{(6)}(0) \rangle}{\langle M(0)^{N_f} \rangle} - 15 \frac{\langle D^{(2)}(0) \rangle \langle D^{(4)}(0) \rangle}{\langle M(0)^{N_f} \rangle^{2}} + 30 \frac{\langle D^{(2)}(0) \rangle^3}{\langle M(0)^{N_f} \rangle^{3}} \bigg{]} \\
&+ {\cal O}\left( \frac{\mu}{T} \right)^8 - \log\left[ \frac{Z}{Z_{YM}} \right] \, .
}
Plugging in the expressions in (\ref{d2}) - (\ref{d4}) leads to
\SP{
&\log \langle e^{2 i p \theta} \rangle_{QCD} \sim \frac{1}{2!} \left( \frac{\mu}{T} \right)^2 \left[ 4p(p+N_f) \langle M(0)^{N_f} \rangle^{-1} \langle M(0)^{N_f - 2} M'(0)^2 \rangle \right] \\
&\hspace{10mm}+ \frac{1}{4!} \left( \frac{\mu}{T} \right)^4 \bigg{[} 16 p^2 (p+N_f)^2 \langle M(0)^{N_f} \rangle^{-1} \langle M(0)^{N_f - 4} M'(0)^4 \rangle \\
&\hspace{10mm}+ 8 p (p+N_f) \langle M(0)^{N_f} \rangle^{-1} \bigg{\langle} M(0)^{N_f - 4} M'(0)\\
&\hspace{10mm}\times \left[ (4 - 3 N_f + N_f^2) M'(0)^3 + 3 (N_f - 2) M(0) M'(0) M''(0) + 2 M(0) M'''(0) \right] \bigg{\rangle}\\
&\hspace{10mm}- 48 p^2 (p+N_f)^2 \langle M(0)^{N_f} \rangle^{-2} \langle M(0)^{N_f - 2} M'(0)^2 \rangle^2 \\
&\hspace{10mm}- 24 p (p+N_f) \langle (N_f - 1) M'(0)^2 + M''(0) M(0) \rangle \langle M(0)^{N_f - 2} M'(0)^2 \rangle \bigg{]} \\
&\hspace{10mm}+{\cal O}\left( \frac{\mu}{T} \right)^6 ~.
}
Therefore, unless some special relationships exist between the expectation values for the terms proportional to $p^2 (p+ N_f)^2$, there are non-Gaussian contributions at ${\cal O}\left( \frac{\mu}{T} \right)^4$. Specifically to obtain a Gaussian form at this order it is required that
\EQ{
\langle M(0)^{N_f} \rangle \langle M(0)^{N_f - 4} M'(0)^4 \rangle = 3 \langle M(0)^{N_f - 2} M'(0)^2 \rangle^2 \, .
\label{taylor-relation-1}
}
At ${\cal O}(\frac{\mu}{T})^6$, it is required, to exclude terms proportional to $p^3 (p+N_f)^3$, that
\SP{
0 ~=~ &\frac{\langle M(0)^{N_f - 6} M'(0)^6 \rangle}{\langle M(0)^{N_f} \rangle} - 15 \frac{\langle M(0)^{N_f - 2} M'(0)^2 \rangle \langle M(0)^{N_f - 4} M'(0)^4 \rangle}{\langle M(0)^{N_f} \rangle^2} \\
&+ 30 \frac{\langle M(0)^{N_f - 2} M'(0)^2 \rangle^3}{\langle M(0)^{N_f} \rangle^3} \, .
\label{taylor-relation-2}
}
There should be an additional requirement to exclude terms proportional to $p^2 (p+N_f)^2$ at this order which we have not computed. Similar requirements are expected at higher orders. We note that these relationships are similar to the ones obtained from a cumulant expansion of the complex phase in \cite{Ejiri:2009hq}.

A calculation of the baryon number moments from the Taylor expansion is provided in Appendix \ref{D}.

\section{Conclusions}
\label{conclude}

    In the context of the reweighting approach to the sign problem, the histogram method combined with a cumulant
expansion \cite{Ejiri:2007ga,Ejiri:2012ng,Ejiri:2012wp,Ejiri:2013lia} is real progress.  Instead of calling for numerical accuracy
of order $e^{-V}$ when the sign problem is severe, the requirement is now reduced to accuracy on the order of some inverse power of the lattice volume $1/V^p$.  The catch, which we have pointed out in this article, is that we do not yet know what that power is.  
 
    If it were really true that the phase angle $\th$ of the fermion determinant follows a Gaussian distribution, then the sign problem is solved.  Only the 2nd order cumulant in the cumulant expansion of $\langle e^{iN_f \th} \rangle_{pq}$ need be computed.  However, as we have shown in sections \ref{sc} and \ref{taylor}, non-Gaussian contributions to the phase angle distribution $\rho_{pq}(\th)$ (the ``density'' in the density of states method) seem inevitable, and grow with hopping parameter, lattice coupling $\b$, and especially chemical potential $\m$.  These contributions lead to non-zero higher-order cumulants, and we see no obvious reason why those cumulants should be negligible in the interesting regions of the phase diagram, where the value of $\m$ is substantial.\footnote{For $\m > \oh m_\pi$ we expect the Gaussian form to break down completely, see e.g.\
\cite{Lombardo:2009aw}.}

    Unfortunately, those higher-order cumulants are not so easy to measure.  By eye, $\rho_{pq}(\th)$ will always look like a
Gaussian; the 4th and higher-order cumulants arise from tiny $O(1/V)$ deviations from the Gaussian form.  As we have seen,
the higher-order cumulants can be significant compared to the 2nd-order cumulant, despite the fact that they arise from 
\emph{apparently} negligible deviations from the Gaussian.  Likewise, it is not sufficient to show that the Binder cumulant $B_4$
is very close to 3.  In order to show that the 4th-order cumulant $(\th^4)_c$ is negligible, one must show that $B_4=3$ to an
accuracy better than $O(1/V)$.     

    The best case scenario is that, for unknown reasons, the higher-order cumulants are negligible even when the chemical
potential is comparatively large.   Our point is that while this best case may be true, it must be shown to be true.
The challenge, then, is to compute the 4th-order cumulant, in interesting regions of the phase diagram, to an accuracy sufficient to show that it is negligible compared to the 2nd-order cumulant.  This requires computing the moments $\langle \th^2 \rangle_{pq}$ and $\langle \th^4 \rangle_{pq}$ to a relative accuracy of at least $O(1/V)$.  If the 4th-order cumulant is non-negligible, then
one must go on to the 6th-order cumulant, which requires computing moments to a relative accuracy of $O(1/V^2)$, and so on
until it is clear that the cumulant expansion has converged.  Numerical accuracies of $O(1/V)$ seem at least feasible, and the optimistic scenario is by no means ruled out.  The worst case scenario is that many higher-order cumulants are required in order to investigate the interesting region of the phase diagram, requiring unattainable levels of numerical accuracy.  Unfortunately this pessimistic scenario is not yet ruled out either.

\begin{acknowledgments}
We would like to thank Jan Rosseel for providing the Mathematica functions used to calculate weights and multiplicities for representations of $SU(n)$. We would like to thank Gert Aarts, Alexander Christensen, Shinji Ejiri, Simon Hands, Tim Hollowood, Jens Langelage, Peter Pedersen, Hana Saito, and participants of the INT program on `Quantum Noise' for discussions. JCM would like to thank the group of Karl Jansen at DESY Zeuthen for hosting her while parts of this work were completed. This work was supported by the U.S. Department of Energy under Grant No. DE-FG03-92ER40711 (JG) and the {\sl Sapere Aude} program of The Danish Council for Independent Research (JCM and KS).
\end{acknowledgments}
\appendix


\section{Example from the hadron resonance gas model}
\label{C}

To calculate the contribution of
\EQ{
\frac{Z_{YM}}{Z} \bigg{\langle} \frac{\det^a {\cal M}(\mu)}{\det^b  {\cal M}(-\mu)} \bigg{\rangle}
\label{gen-mom-ab}
}
in the hadron resonance gas model it is necessary to build up all possible mesons and baryons from $a+b$ quarks, keeping in mind that $b$ are ghost quarks, which contribute an extra factor of $-1$ since we are considering a $1$-loop computation. The partition function is defined by $Z \equiv \langle \det^{a-b}  {\cal M} (\mu) \rangle$.

Consider, for example, the case of $a = 3$, $b = 1$. Since $a = p + N_f$ and $b = p$ in (\ref{phase-moms-hrg}) this corresponds to $p=1$, $N_f=2$.

The baryon contribution is obtained from the $SU(4)$ decomposition,
\EQ{
{\bf 4} \otimes {\bf 4} \otimes {\bf 4} = {\bf 20_{(2)}} \oplus {\bf 4_{(4)}} \oplus {\bf 20_{(2)}} \oplus {\bf 20_{(4)}} \, ,
\label{su4-decomp-baryons}
}
in agreement with (\ref{baryon-decomp}), where the form ${\bf R_{(g)}}$ contains ${\bf R} \in SU(4)$, and ${\bf g} \in SU(2)$ obtained from the $SU(8)$ decomposition of the symmetric representation in (\ref{su2n-to-sun}), with help of the identity constraint at $\mu = 0$.

The weights of the above representations of $SU(4)$ are given in Table \ref{weight-tab}. Note that in (\ref{su4-decomp-baryons}) there are two ${\bf 20}$'s for each ${\bf 4_A}$ and ${\bf 20_S}$. Keeping this in mind, the right-hand column gives the total multiplicity and the quark content of each weight, where each weight can be rewritten as a sum of three of the fundamental weights, $q_1 \equiv (1,0,0)$, $q_2 \equiv (-1,1,0)$, $q_3 \equiv (0,-1,1)$, and $q_4 \equiv (0,0,-1)$.

\begin{table}[h]
\begin{tabular}{|l|l|l|l|}
\hline
${\bf 20}$ & ${\bf 4_A}$ & ${\bf 20_{S}}$ & baryon contribution\\
\hline
&& $(3,0,0)$ & $[q_1 q_1 q_1]$\\
\hline
$(1,1,0)$ && $(1,1,0)$ & $3 \times [q_1 q_1 q_2]$\\
\hline
$(-1,2,0)$ && $(-1,2,0)$ & $3 \times [q_1 q_2 q_2]$\\
\hline
$(2,-1,1)$ && $(2,-1,1)$ & $3 \times [q_1 q_1 q_3]$\\
\hline
&& $(-3,3,0)$ & $[q_2 q_2 q_2]$\\
\hline
$2 \times (0,0,1)$ & $(0,0,1)$ & $(0,0,1)$ & $6 \times [q_1 q_2 q_3]$\\
\hline
$(2,0,-1)$ && $(2,0,-1)$ & $3 \times [q_1 q_1 q_4]$\\
\hline
$(-2,1,1)$ && $(-2,1,1)$ & $3 \times [q_2 q_2 q_3]$\\
\hline
$(1,-2,2)$ && $(1,-2,2)$ & $3 \times [q_1 q_3 q_3]$\\
\hline
$2 \times (0,1,-1)$ & $(0,1,-1)$ & $(0,1,-1)$ & $6 \times [q_1 q_2 q_4]$\\
\hline
$(-1,-1,2)$ && $(-1,-1,2)$ & $3 \times [q_2 q_3 q_3]$\\
\hline
$(-2,2,-1)$ && $(-2,2,-1)$ & $3 \times [q_2 q_2 q_4]$\\
\hline
$2 \times (1,-1,0)$ & $(1,-1,0)$ & $(1,-1,0)$ & $6 \times [q_1 q_3 q_4]$\\
\hline
&& $(0,-3,3)$ & $[q_3 q_3 q_3]$\\
\hline
$2 \times (-1,0,0)$ & $(-1,0,0)$ & $(-1,0,0)$ & $6 \times [q_2 q_3 q_4]$\\
\hline
$(1,0,-2)$ && $(1,0,-2)$ & $3 \times [q_1 q_4 q_4]$\\
\hline
$(0,-2,1)$ && $(0,-2,1)$ & $3 \times [q_3 q_3 q_4]$\\
\hline
$(-1,1,-2)$ && $(-1,1,-2)$ & $3 \times [q_2 q_4 q_4]$\\
\hline
$(0,-1,-1)$ && $(0,-1,-1)$ & $3 \times [q_3 q_4 q_4]$\\
\hline
&& $(0,0,-3)$ & $[q_4 q_4 q_4]$\\
\hline
\end{tabular}
\caption{Weights of representations of the $SU(4)$ given in (\ref{su4-decomp-baryons}) and baryon configurations.}
\label{weight-tab}
\end{table}

Notice that each weight with a multiplicity of $3$ comes from two baryons with $g = 2$, and one with $g = 4$. When the multiplicity is $6$ there are four contributions from $g = 2$ and two from $g = 4$. Each weight with a multiplicity $1$ just gives a contribution with $g = 4$. If we take $q_4$ to represent the $\det  {\cal M}(-\mu)$ in the denominator of (\ref{gen-mom-ab}), then each contribution of $q_4$ gives a multiplicative factor of $-1$ to the free energy and adds a contribution of $-\mu$ to the baryon chemical potential.

Since we are considering a $1$-loop computation, the hadrons are non-interacting and the free energies can simply be summed. The contribution from the weights of total multiplicity $1$ is
\EQ{
\left[ 2 G_4(3\mu) \right] \, ,
}
for total multiplicity $3$,
\EQ{
 2 \left[ 6 G_{2}(3\mu) \right] + \left[ 6 G_4(3\mu) \right] \, ,
 }
and for total multiplicity $6$,
 \EQ{
 4 \left[ G_2(3\mu) - 3 G_2(\mu) \right] + 2 \left[ G_4(3\mu) - 3 G_4(\mu) \right] \, .
}
Adding these up gives
\EQ{
10 G_4(3\mu) - 6 G_4(\mu) + 16 G_2(3\mu) - 12 G_2(\mu) \, .
\label{free-E-no-Z}
}
Now all that's left is the contribution from the partition function in the denominator of (\ref{gen-mom-ab}). Since $a-b=2$ in this example the contribution from the partition function is obtained from the $SU(2)$ decomposition
\EQ{
{\bf 2} \otimes {\bf 2} \otimes {\bf 2} = {\bf 2}_{(2)} \oplus {\bf 2}_{(2)} \oplus {\bf 4}_{(4)} \, ,
\label{su2-decomp}
}
where the subscripts ($g$ values) are determined using (\ref{su2n-to-sun}). Since there are no ghost quarks in the partition function, there are only contributions from $G_2(3\mu)$ and $G_4(3\mu)$. Therefore using (\ref{su2-decomp}), the contribution to the free energy is
\EQ{
-4 G_4(3\mu) - 4 G_2(3\mu) \, .
}
Adding this result to the free energy in (\ref{free-E-no-Z}) gives the total contribution, up to an overall constant multiplicative factor,
\EQ{
6 [ G_4(3\mu) - G_4(\mu) ] + 12 [ G_2(3\mu) - G_2(\mu) ] \, .
}
Since we're considering $p=1$, $N_f=2$, $p(p+N_f) = 3$, this result fits the form $p(p+N_f)X_1$.

The meson contribution is obtained from the $SU(8)$ decomposition (\ref{meson-decomp}),
\EQ{
{\bf 8} \otimes {\bf {\bar 8}} = {\bf 15_{3}} \oplus {\bf 15_{1}} \oplus {\bf 1_{3}} \oplus {\bf 1_{1}} \, ,
\label{su8-decomp-mesons}
}
where ${\bf R_g}$ contains ${\bf R} \in SU(4)$, and $g \in SU(2)$. Note that both the ${\bf 15}$ and ${\bf 1}$ of $SU(4)$ have an equal probability of having $g = 1$ or $g = 3$. All quark anti-quark combinations are equally likely given that the determinants are identical with the exception of $\mu$, all we need to know are the combinations which are listed in Table \ref{meson-configs}. Note that the weights of the anti-fundamental are ${\bar q_1} \equiv (-1,0,0)$, ${\bar q_2} \equiv (1,-1,0)$, ${\bar q_3} \equiv (0,1,-1)$, and ${\bar q_4} \equiv (0,0,1)$.

\begin{table}[h]
\begin{tabular}{|l|l|l|}
\hline
${\bf 0}$ & ${\bf 15}$ & meson contribution\\
\hline
& $(1,0,1)$ & $[q_1 {\bar q_4}]$\\
\hline
& $(-1,1,1)$ & $[q_2 {\bar q_4}]$\\
\hline
& $(1,1,-1)$ & $[q_1 {\bar q_3}]$\\
\hline
& $(0,-1,2)$ & $[q_3 {\bar q_4}]$\\
\hline
& $(-1,2,-1)$ & $[q_2 {\bar q_3}]$\\
\hline
& $(2,-1,0)$ & $[q_1 {\bar q_2}]$\\
\hline
$(0,0,0)$ & $3 \times (0,0,0)$ & $[q_1 {\bar q_1}] + [q_2 {\bar q_2}] + [q_3 {\bar q_3}] + [q_4 {\bar q_4}]$\\
\hline
& $(-2,1,0)$ & $[q_2 {\bar q_1}]$\\
\hline
& $(1,-2,1)$ & $[q_3 {\bar q_2}]$\\
\hline
& $(0,1,-2)$ & $[q_4 {\bar q_3}]$\\
\hline
& $(-1,-1,1)$ & $[q_3 {\bar q_1}]$\\
\hline
& $(1,-1,-1)$ & $[q_4 {\bar q_2}]$\\
\hline
& $(-1,0,-1)$ & $[q_4 {\bar q_1}]$\\
\hline
\end{tabular}
\caption{Weights of representations of the $SU(4)$ given in (\ref{su8-decomp-mesons}) and meson configurations.}
\label{meson-configs}
\end{table}

The contribution to the expectation value in (\ref{gen-mom-ab}) from the mesons is then
\EQ{
4 [G_1(0) + G_3(0)] + 6 [G_1(0) + G_3(0)] - 6 [G_1(2\mu) + G_3(2\mu)] \, .
}
The partition function contributes
\EQ{
-4 G_1(0) - 4 G_3(0) \, ,
}
so the total meson contribution is, up to an overall multiplicative factor,
\EQ{
6 [G_1(0) - G_1(2\mu)] + 6 [G_3(0) - G_3(2\mu)] \, .
}
Since $p(p+N_f)=3$, this also fits the form $p(p+N_f)X_1$.
\section{Distribution of the baryon number}
\label{D}

The calculation of the baryon number distribution proceeds in much the same way as that for the complex phase in 
Section \ref{pad}. It is obtained via the Fourier transform
\EQ{
\langle n_{B} \delta(\theta - \theta') \rangle_{QCD} = 2 \int_{-\infty}^{\infty} \frac{{\rm d}p}{2 \pi} e^{-2 i p \theta'} \langle n_{B} e^{2 i p \theta} \rangle_{QCD} \, ,
\label{baryon-num-dist}
}
where the moments are naturally given by
\EQ{
\langle n_{B} e^{2 i p \theta} \rangle_{QCD} = \lim_{{\tilde \mu} \rightarrow \mu} \frac{\partial}{\partial {\tilde \mu}} \bigg{\langle} \frac{\det^{p} (\Dsl + \gamma_0 \mu + m)}{\det^{p} (\Dsl - \gamma_0 \mu + m)} {\det}^{N_f} (\Dsl + \gamma_0 {\tilde \mu} + m) \bigg{\rangle} \frac{Z_{YM}}{Z} \, .
\label{baryon-num-moms}
}
In what follows we calculate this quantity from the strong coupling expansion and the hadron resonance gas model, and also consider the Taylor expansion. Our results for the expectation values $\langle \frac{M(\mu)^p}{M(-\mu)^p} M({\tilde \mu})^{N_f} \rangle \frac{Z_{YM}}{Z}$, with $M(\mu) \equiv \det (\Dsl + \gamma_0 \mu + m)$, take the form of an exponential of a polynomial in $p$,
\EQ{
\bigg{\langle} \frac{\det^{p} (\Dsl + \gamma_0 \mu + m)}{\det^{p} (\Dsl - \gamma_0 \mu + m)} {\det}^{N_f} (\Dsl + \gamma_0 {\tilde \mu} + m) \bigg{\rangle} \frac{Z_{YM}}{Z} = e^{k_0 + k_1 p + k_2 p^2 + ...} \, .
\label{baryon-num-moms-gen}
}
Therefore, taking the derivative and the limit in (\ref{baryon-num-moms}) puts the baryon number moments in the form
\SP{
\langle n_{B} e^{2 i p \theta} \rangle_{QCD} &= (c_0 + c_1 p + c_2 p^2 + ...) \langle e^{2 i p \theta} \rangle_{QCD} \\
&= (c_0 + c_1 p + c_2 p^2 + ...) e^{-p(p+N_f)X_1- p^2(p+N_f)^2 X_2 - p^3 (p+N_f)^3 X_3 - ...} \, ,
\label{gen-baryon-moms}
}
where the constants $c_j$ are the derivatives of the $k_j$ in (\ref{baryon-num-moms-gen}), $c_j\equiv\lim_{\tilde\mu\to\mu}\partial_{\tilde \mu}k_j$.

In the case of the hadron resonance gas model, and the large $N_c$ strong coupling expansion, our results fit the simplified form
\EQ{
\langle n_{B} e^{2 i p \theta} \rangle_{QCD} = (c_0 + c_1 p + c_2 p^2) e^{-p(p+N_f)X_1} \, ,
\label{baryon-num-moms-result}
}
where the $c_i$ depend on which model is considered. Performing the integral in (\ref{baryon-num-dist}) then gives the baryon number distribution which can be used to obtain the baryon number by integrating over the phase angle
\EQ{
\langle n_B \rangle_{QCD} = \int_{-\infty}^{\infty} {\rm d}\theta ~ \langle n_B \delta(\theta - \theta') \rangle_{QCD} \, .
}
Again, when considering the strong coupling expansion for $N_c = 3$ there are corrections to the result in (\ref{baryon-num-moms-result}). In this case the moments take the more general form in (\ref{gen-baryon-moms}). Regardless of whether the moments take the form in (\ref{gen-baryon-moms}) or in (\ref{baryon-num-moms-result}) we found in \cite{Greensite:2013vza} that the entire contribution to the average baryon number comes from the $c_0$ term and the contributions from the terms with $c_1$, $c_2$, ... constitute the noise.

It is interesting in general how the signal of distributions in systems with a sign problem is separated from the noise. In a recent analysis of data from a unitary fermi gas the signal for the ground state energies was found to be well described by the first few terms in a cumulant expansion of the log of the relevant correlator \cite{Endres:2011jm,Endres:2011mm,Endres:2012cw}.

\subsection{Distribution of the baryon number in a free hadron gas}

The baryon number distribution can be calculated from the moments in (\ref{baryon-num-moms}) by means of a straightforward generalization of the procedure in Section \ref{hrg}. The quantity of interest is the expectation value
\EQ{
\bigg{\langle} \frac{\det^{p} (\Dsl + \gamma_0 \mu + m)}{\det^{p} (\Dsl - \gamma_0 \mu + m)} {\det}^{N_f} (\Dsl + \gamma_0 {\tilde \mu} + m) \bigg{\rangle} \frac{Z_{YM}}{Z} = \alpha_M(\mu,{\tilde \mu}) \alpha_B(\mu,{\tilde \mu}) \, ,
}
where $\alpha_M$ is the meson contribution and $\alpha_B$ is the baryon contribution. The possible combinations are again determined using the group theory methods in the Section \ref{hrg} and Appendix \ref{C}. The results for the combinatorics are similar. For mesons, accounting for the quark content of each of the weights and taking into account the multiplicities gives rise to the structure $\left[ (p q_3+N_f q_1-p q_2)(p{\bar q_3}+N_f {\bar q_1}-p{\bar q_2}) - N_f^2 q_3{\bar q_3} \right]$, while for baryons one obtains $-\left[ (p q_3+N_f q_1-p q_2)^3 - N_f^3 q_3^3 \right]$. The quark content $q_1$ corresponds to one of the $p$ determinants in the numerator with chemical potential $\mu$, $q_3$ corresponds to one of the $p$ determinants in the denominator with chemical potential $-\mu$, $q_3$ corresponds to one of the $N_f$ determinants with chemical potential ${\tilde \mu}$. The meson contribution thus takes the form
\SP{
\alpha_{M}(\mu,{\tilde \mu}) = \exp \bigg[ &p^2 \left( G^{M{\bar M}}(0) - G^{M{\bar M}}(2\mu) \right) + N_f p \left( -G^{M{\bar M}}({\tilde \mu},-\mu) + G^{M{\bar M}}({\tilde \mu},\mu) \right) \bigg] \, ,
}
in agreement with \cite{Lombardo:2009aw}. 
The baryon contribution is given by
\SP{
\alpha_{B}(\mu,{\tilde \mu}) &= \exp \bigg[ N_f p^2 \left(-G^{B{\bar B}}({\tilde \mu},-\mu,-\mu) + 2 G^{B{\bar B}}({\tilde \mu},\mu,-\mu) - G^{B{\bar B}}({\tilde \mu},\mu,\mu) \right)\\
&+ N_f^2 p \left( G^{B{\bar B}}({\tilde \mu},{\tilde \mu},-\mu) - G^{B{\bar B}}({\tilde \mu},{\tilde \mu},\mu) \right) + \left( - {\tilde G}^{B{\bar B}}({\tilde \mu},{\tilde \mu},{\tilde \mu}) + {\tilde G}^{B{\bar B}}(\mu,\mu,\mu) \right) \bigg] \, ,
}
with
\EQ{
{\tilde G}^{B{\bar B}}(\mu_1,\mu_2,\mu_3) \equiv {\tilde k}_B \left[ \frac{2}{3} N_f (N_f^2 - 1) G_2^{B{\bar B}}(\mu_1,\mu_2,\mu_3) + \frac{1}{3} N_f (N_f^2 + 2) G_4^{B{\bar B}}(\mu_1,\mu_2,\mu_3) \right]  \, ,
}
where the first term gives the number of $g=2$ states in ${\bf N_f} \otimes {\bf N_f} \otimes {\bf N_f}$ and the second term gives the number of $g = 4$ states, using (\ref{baryon-decomp}).

Notice that for ${\tilde \mu} \rightarrow \mu$, $\alpha_M(\mu,{\tilde \mu}) \rightarrow \alpha_M(\mu)$ in (\ref{alphaM}) and $\alpha_B(\mu,{\tilde \mu}) \rightarrow \alpha_B(\mu)$ in (\ref{alphaB}) such that $\alpha_M(\mu,{\tilde \mu}) \alpha_M(\mu,{\tilde \mu}) \rightarrow e^{-p(p+N_f)X_1}$ in (\ref{complex-phase-moms-result}). It is now possible to compute the baryon number moments,
\SP{
&\langle n_B e^{2 i p \theta} \rangle = \lim_{{\tilde \mu} \rightarrow \mu} \frac{\partial}{\partial {\tilde \mu}} \bigg{\langle} \frac{\det^{p} (\Dsl + \gamma_0 \mu + m)}{\det^{p} (\Dsl - \gamma_0 \mu + m)} {\det}^{N_f} (\Dsl + \gamma_0 {\tilde \mu} + m) \bigg\rangle \frac{Z_{YM}}{Z} \\
&= \left( c_0 + c_1 p + c_2 p^2 \right) e^{-p(p+N_f)X_1} \, ,
}
where $X_1$ is given in (\ref{X-def}) and
\EQ{
c_0 \equiv - \frac{\partial}{\partial {\tilde \mu}} {\tilde G}^{B{\bar B}}({\tilde \mu},{\tilde \mu},{\tilde \mu}) \bigg{|}_{{\tilde \mu} \rightarrow \mu} \, ,
}
\SP{
c_1 \equiv &\frac{\partial}{\partial {\tilde \mu}} N_f \left[ -G^{M{\bar M}}({\tilde \mu},-\mu) + G^{M{\bar M}}({\tilde \mu},\mu) \right]\\
&+ \frac{\partial}{\partial {\tilde \mu}} N_f^2 \left[ G^{B{\bar B}}({\tilde \mu},{\tilde \mu},-\mu) - G^{B{\bar B}}({\tilde \mu},{\tilde \mu},\mu) \right] \bigg{|}_{{\tilde \mu} \rightarrow \mu} \, ,
}
\EQ{
c_2 \equiv \frac{\partial}{\partial {\tilde \mu}} N_f \left[ -G^{B{\bar B}}({\tilde \mu},-\mu,-\mu) + 2 G^{B{\bar B}}({\tilde \mu},\mu,-\mu) - G^{B{\bar B}}({\tilde \mu},\mu,\mu) \right] \bigg{|}_{{\tilde \mu} \rightarrow \mu} \, .
}

\subsection{Baryon number moments from the strong coupling expansion}

To obtain $\langle n_B e^{2 i p \theta} \rangle_{QCD}$ in (\ref{baryon-num-moms}) from the combined strong coupling and hopping expansion it is necessary to solve for the expectation value
\SP{
{\tilde Q} &\equiv \bigg{\langle} \frac{\det^p (\Dsl + \gamma_0 \mu + m)}{\det^p (\Dsl - \gamma_0 \mu + m)} \det^{N_f} (\Dsl + \gamma_0 {\tilde \mu} + m) \bigg{\rangle} \\
&= 1 + {\tilde q}_1 h + {\tilde q}_2 h^2 + {\tilde q}_3 h^3 + ... \, ,
\label{Q-tilde-expansion}
}
which is analogous to (\ref{Q-expansion}) except for the extra ${\tilde \mu}$-dependence. We again use the action in (\ref{ferm-action}) to obtain the expectation values.

\subsubsection{${\cal O}(h^2)$}

Expanding ${\tilde Q}$ to ${\cal O}(h^2)$ using (\ref{ferm-action}) one obtains
\EQ{
{\tilde q}_1 = a_1 \left[ p (e^{\mu/T} - e^{-\mu/T}) \left( \langle L_1 \rangle - \langle L_1^* \rangle \right) + N_f (e^{{\tilde \mu}/T} \langle L_1 \rangle + e^{-{\tilde \mu}/T} \langle L_1^* \rangle ) \right] \, ,
}
\SP{
{\tilde q}_2 = ~&\frac{1}{2} a_1^2 p^2 (e^{\mu/T} - e^{-\mu/T})^2 \langle (L_1 - L_1^*)^2 \rangle \\
&+ p N_f a_1^2 \big{[} (e^{\mu/T+{\tilde \mu}/T} - e^{-\mu/T+{\tilde \mu}/T}) \langle L_1^2 \rangle - (e^{\mu/T} - e^{-\mu/T})(e^{{\tilde \mu}/T} - e^{-{\tilde \mu}/T}) \langle L_1 L_1^* \rangle \\
&+ (e^{-\mu/T-{\tilde \mu}/T} - e^{\mu/T - {\tilde \mu}/T}) \langle (L_1^*)^2 \rangle \big{]} +a_2 p(e^{2\mu/T} - e^{-2\mu/T}) \langle L_2 - L_2^* \rangle \\
&+ a_2 N_f (e^{2 {\tilde \mu}/T} \langle L_2 \rangle + e^{-2{\tilde \mu}/T} \langle L_2^* \rangle) + \frac{1}{2} a_1^2 N_f^2 \langle (e^{{\tilde \mu}/T}L_1 + e^{-{\tilde \mu}/T}L_1^*)^2 \rangle \, ,
}
where all expectation values are with respect to the Yang-Mills vacuum. Therefore $\langle L_n^j \rangle = \langle (L_n^*)^j \rangle$. The ${\tilde q}_i$ then simplify to
\EQ{
{\tilde q}_1 = 2 a_1 N_f \cosh({\tilde \mu}/T) \langle L_1 \rangle \, ,
}
\SP{
{\tilde q}_2 = &2 a_1^2 p^2 \left[ \cosh(2\mu/T) - 1 \right] \left[ \langle L_1^2 \rangle - \langle L_1 L_1^* \rangle \right] \\
&+ 2 a_1^2 p N_f \left[ \cosh(\mu/T + {\tilde \mu}/T) - \cosh(\mu/T - {\tilde \mu}/T) \right] \left[ \langle L_1^2 \rangle - \langle L_1 L_1^* \rangle \right] \\
&+ 2 a_2 N_f \cosh(2 {\tilde \mu}/T) \langle L_2 \rangle + a_1^2 N_f^2 \left[ \cosh(2{\tilde \mu}/T) \langle L_1^2 \rangle + \langle L_1 L_1^* \rangle \right] \, .
}
Note that the ${\tilde q}_i$ reduce to the $q_i$ in (\ref{q1h2}) and (\ref{q2h2}) when ${\tilde \mu} \rightarrow \mu$. In the confined phase $\langle L_1 \rangle = \langle L_1^2 \rangle = 0$, $\langle L_1 L_1^* \rangle = N_s$, and $\langle L_2 \rangle= \sum_x \langle (\tr W_x)^2 - 2~ \tr_{AS} W_x \rangle = 0$ (AS refers to the antisymmetric representation), up to ${\cal O}(\lambda_1)$. In this case
\EQ{
{\tilde q}_1 = 0 \, ,
}
\EQ{
{\tilde q}_2 = -4 a_1^2 p^2 N_s \sinh^2(\mu/T) - 4 a_1^2 p N_s N_f \sinh(\mu/T) \sinh({\tilde \mu}/T) + a_1^2 N_s N_f^2 \, .
\label{tilde-q2}
}
Since the only ${\tilde \mu}$-dependent contribution is at ${\cal O}(p)$, this result will only give us the leading contribution to $c_1$ when we take the derivative with respect to ${\tilde \mu}$ in (\ref{baryon-num-moms}).

\subsubsection{${\cal O}(h^3)$}

In order to obtain the leading contribution to $c_0$ it is necessary to calculate the ${\cal O}(p^0)$ contribution to ${\tilde q}_3$ in (\ref{Q-tilde-expansion}), which takes the form
\SP{
{\tilde q}_3 = \bigg[ &\left( 2 a_1 a_2 N_f^2 \langle L_2 L_1^* \rangle + a_1^3 N_f^3 \langle L_1^2 L_1^* \rangle \right) \cosh({\tilde \mu}/T) \\
&+ \left( 2 a_3 N_f \langle L_3 \rangle + 2 a_1 a_2 N_f^2 \langle L_1 L_2 \rangle + \frac{1}{3} a_1^3 N_f^3 \langle L_1^3 \rangle \right) \cosh(3{\tilde \mu}/T) \bigg] + {\cal O}(p) + {\cal O}(p^2) \, .
\label{tilde-q3-init}
}
In the confined phase it is possible to work out the expectation values using (\ref{group-int}) along with the Frobenius formula to convert $L_n$ with $n > 1$ into combinations of $L_{R}$, where $R$ are representations of $SU(N_c)$ (see for example Appendix B of \cite{Myers:2009df}). The expectation values which have color singlets in their group decompositions contribute. One can check that, for $SU(3)$, up to ${\cal O}(\lambda_1)$,
\SP{
\langle L_2 L_1^* \rangle &= 0 \, , \\
\langle L_1^2 L_1^* \rangle &= 0 \, , \\
\langle L_1^3 \rangle &= N_s \, , \\
\langle L_3 \rangle &= N_s \, , \\
\langle L_1 L_2 \rangle &= - N_s \, .
\label{su3-mult-wind}
}
To obtain the above we used, for $SU(3)$, the following identities from the Frobenius formula: $\tr (W^2) = (\tr W)^2 - 2~ \tr W^{\dagger}$, and $2~ \tr (W^3) = 6~ \tr_{{\bf 10}_S} W - (\tr W)^3 - 3~ \tr W \tr (W^2)$, where ${\bf 10}_S$ refers to the symmetric representation of $SU(3)$ with dimension $10$. Emphasizing the ${\cal O}(p^0)$ contribution, the result for ${\tilde q}_3$ in the confined phase is then
\SP{
{\tilde q}_3 = N_s \bigg[ 2 a_3 N_f - 2 a_1 a_2 N_f^2 + \frac{1}{3} a_1^3 N_f^3 \bigg] \cosh(3{\tilde \mu}/T) + {\cal O}(p) + {\cal O}(p^2) \, .
\label{tilde-q3-p0}
}

The leading order contribution to $c_1$ was obtained at ${\cal O}(h^2)$ from ${\tilde q_2}$, however, the contribution at ${\cal O}(p^2)$ was independent of ${\tilde \mu}$. Therefore the leading order contribution to $c_2$ should also occur at ${\cal O}(h^3)$. We rewrite (\ref{tilde-q3-init}) as
\SP{
{\tilde q}_3 = 4 p^2 \bigg{[} &4 a_1 a_2 \left[ \langle L_1 L_2 \rangle - \langle L_2 L_1^* \rangle \right] \cosh(\mu/T) \\
&+ a_1^3 N_f \left[ \langle L_1^3 \rangle - \langle L_1^2 L_1^* \rangle \right] \cosh({\tilde \mu}/T) \bigg{]} \sinh^2(\mu/T) + {\cal O}(p) + {\cal O}(p^0) \, ,
}
emphasizing the contribution at ${\cal O}(p^2)$. Using the relations in (\ref{su3-mult-wind}) for the confined phase, and working to ${\cal O}(\lambda_1)$, this result simplifies to
\EQ{
{\tilde q}_3 = 4 N_s p^2 \bigg{[} &-4 a_1 a_2 \cosh(\mu/T) + a_1^3 N_f \cosh({\tilde \mu}/T) \bigg{]} \sinh^2(\mu/T) + {\cal O}(p) + {\cal O}(p^0) \, .
\label{tilde-q3-p2}
}

\subsubsection{${\cal O}(h^4)$ and summary of results}

To obtain $c_3$, considering the ${\cal O}(h^4)$ contribution to ${\tilde Q}$, keeping the ${\cal O}(p^3)$ and ${\cal O}(p^4)$ contributions. We obtain
\SP{
{\tilde q}_4 = &\frac{4}{3}a_1^4 \left[ p^4 \sinh^4(\mu/T) + 2 p^3 N_f \sinh^3(\mu/T) \sinh({\tilde \mu}/T) \right] \left[ \langle L_1^4 \rangle - 4 \langle L_1^3 L_1^* \rangle + 3 \langle L_1^2 L_1^{* 2} \rangle \right] \\
&+ {\cal O}(p^2) \, .
}
This reduces to (\ref{q4}) in the limit ${\tilde \mu} \rightarrow \mu$. The expectation values were are worked out in the confined phase in Appendix \ref{E} to obtain
\EQ{
\langle L_1^4 \rangle - 4 \langle L_1^3 L_1^* \rangle + 3 \langle L_1^2 L_1^{* 2} \rangle = 3 [2 N_s^2 + 6 N_s \lambda_1 + 24 N_s^2 \lambda_1] + {\cal O}(\lambda_1^2) \, .
}
Therefore the contribution to ${\tilde q}_4$ up to ${\cal O}(\lambda_1)$ in the confined phase is
\SP{
{\tilde q}_4 = &4 a_1^4 \left[ p^4 \sinh^4(\mu/T) + 2 p^3 N_f \sinh^3(\mu/T) \sinh({\tilde \mu}/T) \right] \left[2 N_s^2 + 6 N_s \lambda_1 + 24 N_s^2 \lambda_1 \right] \\
&+ {\cal O}(p^2) \, .
\label{tilde-q4-p3}
}
Note that the presence of the ${\cal O}(V)$ contribution (the term proportional to $N_s$), at ${\cal O}(p^3)$ and ${\cal O}(p^4)$ supports the more general form for the baryon number moments in (\ref{gen-baryon-moms}),
\SP{
\langle n_{B} e^{2 i p \theta'} \rangle_{QCD} &= (c_0 + c_1 p + c_2 p^2 + ...) e^{-p(p+N_f)X_1- p^2(p+N_f)^2 X_2 - ...} \, ,
}
where the $c_i$ include the derivatives with respect to $\tilde \mu$ in (\ref{baryon-num-moms}), followed by taking the limit ${\tilde \mu} \rightarrow \mu$. The leading contributions in the confined phase up to ${\cal O}(\lambda_1)$ are given by
\SP{
c_0 &= \frac{N_s}{T} h^3 \left[ 6 a_3 N_f  - 6 a_1 a_2 N_f^2 + a_1^3 N_f^3 \right] \sinh(3 \mu/T) + {\cal O}(h^4) \, , \\
c_1 &= - \frac{4}{T} a_1^2 h^2 N_s N_f \sinh(\mu/T) \cosh(\mu/T) + {\cal O}(h^3) \, , \\
c_2 &= \frac{4}{T} a_1^3 h^3 N_s N_f \sinh^3 (\mu/T) + {\cal O}(h^4) \, , \\
c_3 &= \frac{16}{T} a_1^4 h^4 N_f \sinh^3(\mu/T) \cosh(\mu/T) \left[ N_s^2 + 3 N_s \lambda_1 + 12 N_s^2 \lambda_1 \right] + {\cal O}(h^5) \, , \\
&... \, ,
}
which are obtained from (\ref{tilde-q3-p0}), (\ref{tilde-q2}), (\ref{tilde-q3-p2}), and (\ref{tilde-q4-p3}), respectively.

\subsection{Baryon number moments from the Taylor expansion}

The baryon number moments $\langle n_B e^{2 i p \theta} \rangle_{QCD}$ can be obtained from the Taylor expansion in a similar manner, using the simplified expression
\SP{
\langle n_B e^{2 i p \theta} \rangle_{QCD} &= \lim_{{\tilde \mu} \rightarrow \mu} \frac{\partial}{\partial {\tilde \mu}} \bigg{\langle} \frac{M(\mu)^p}{M(-\mu)^p} M({\tilde \mu})^{N_f} \bigg{\rangle} \frac{Z_{YM}}{Z} \\
&= N_f \bigg{\langle} \frac{M(\mu)^{p + N_f - 1}}{M(-\mu)^{p}} M'(\mu) \bigg{\rangle} \frac{1}{\langle M(\mu)^{N_f} \rangle} \, .
}
Since the quantity $\langle \frac{M(\mu)^p}{M(-\mu)^p} M({\tilde \mu})^{N_f} \rangle$ is given by an exponential the baryon number moments will necessarily take the form
\SP{
\langle n_{B} e^{2 i p \theta} \rangle_{QCD} &= \left[ c_0 + c_1 p + c_2 p^2 + ... \right] \frac{Z_{YM}}{Z} \bigg{\langle} \frac{M(\mu)^{p+N_f}}{M(-\mu)^p} \bigg{\rangle} \\
&= \left[ c_0 + c_1 p + c_2 p^2 + ... \right] \langle e^{2 i p \theta} \rangle_{QCD} \, .
}
Therefore, to determine the values corresponding to $c_0$, $c_1$, ..., consider the Taylor expansion of the ratio
\SP{
&\frac{\langle n_B e^{2 i p \theta} \rangle_{QCD}}{\langle e^{2 i p \theta} \rangle_{QCD}} = \left( \frac{\mu}{T} \right) N_f \frac{\left( 2 p + N_f - 1 \right) \langle M(0)^{N_f - 2} M'(0)^2 \rangle + \langle M(0)^{N_f - 1} M''(0) \rangle}{\langle M(0)^{N_f} \rangle} \\
&\hspace{3mm}+ \left( \frac{\mu}{T} \right)^3 N_f \bigg{[} \frac{4}{3} \left( \frac{\langle M(0)^{N_f - 4} M'(0)^4 \rangle}{\langle M(0)^{N_f} \rangle} + 3 \langle M(0)^{N_f - 2} M'(0)^2 \rangle \frac{\langle M'(0)^2 \rangle}{\langle M(0)^{N_f} \rangle^{2}} \right) p^3 \\
&\hspace{26mm}+ {\cal O}(p^2) \bigg{]} \\
&\hspace{3mm}+ \left( \frac{\mu}{T} \right)^5 N_f \bigg{[} \frac{4}{15} \bigg{(} \frac{\langle M(0)^{N_f - 6} M'(0)^6 \rangle}{\langle M(0)^{N_f} \rangle} + 10 \langle M(0)^{N_f - 4} M'(0)^{4} \rangle \frac{\langle M'(0)^2 \rangle}{\langle M(0)^{N_f} \rangle^2} \\
&\hspace{35mm}+ 5 \langle M(0)^{N_f - 2} M'(0)^2 \rangle \frac{\langle M'(0)^4 \rangle}{\langle M(0)^{N_f} \rangle^4} \bigg{)} p^5 + {\cal O}(p^4) \bigg{]} \\
&\hspace{3mm}+ {\cal O}\left( \frac{\mu}{T} \right)^7 \, .
}
The baryon number is given by the $p$-independent contributions. Therefore, the $p$-dependent contributions are all expected to vanish upon integration of $e^{-2 i p \theta'} \langle n_B e^{2 i p \theta} \rangle_{QCD}$ over $p$ and $\theta'$.

\section{${\cal O}(h^4)$ corrections from the strong coupling expansion}
\label{E}

To determine if the form $Q \frac{Z_{YM}}{Z} = \exp[-p(p+N_f)X_1]$ can be obtained from the strong coupling action it is helpful to consider the ${\cal O}(h^4 p^3)$ and ${\cal O}(h^4 p^4)$ contributions to the expansion of $Q$ in (\ref{Q-expansion}),
\EQ{
q_4 h^4 = \frac{4}{3}a_1^4 h^4 (p^4 + 2 p^3 N_f) \sinh^4(\mu/T) \left[ \langle L_1^4 \rangle - 4 \langle L_1^3 L_1^* \rangle + 3 \langle L_1^2 L_1^{* 2} \rangle \right] + {\cal O}(p^2) \, .
\label{q4h4}
}
To obtain the expected exponentiation this expression would need to be equivalent to $\frac{1}{2} \left[ p(p+N_f) X_1 \right]^2$, with $X_1 = 2 a_1^2 h^2 \left[ \cosh(2\mu/T) - 1\right] \left[ \langle L_1 L_1^* \rangle - \langle L_1^2 \rangle \right]$, as suggested by our results for $q_2$, up to ${\cal O}(p^2)$ contributions. Plugging in $X_1$,
\SP{
\frac{1}{2} \left[ p(p+N_f) X_1 \right]^2 = &8 a_1^4 h^4 (p^4 + 2 p^3 N_f) \sinh^4(\mu/T) \left[ \langle L_1 L_1^* \rangle^2 - 2 \langle L_1 L_1^* \rangle \langle L_1^2 \rangle + \langle L_1^2 \rangle^2 \right] \\
&+ {\cal O}(p^2) \, .
\label{halfppnfx}
}
To get the exponentiation which would lead to a Gaussian distribution, (\ref{q4h4}) needs to be equal to (\ref{halfppnfx}), up to ${\cal O}(p^2)$ corrections, resulting in the requirement
\EQ{
\langle L_1^4 \rangle - 4 \langle L_1^3 L_1^* \rangle + 3 \langle L_1^2 L_1^{* 2} \rangle = 6 \left[ \langle L_1 L_1^* \rangle^2 - 2 \langle L_1 L_1^* \rangle \langle L_1^2 \rangle + \langle L_1^2 \rangle^2 \right] \, .
\label{special-relation}
}
Is this true?

First consider the contributions from the l.h.s. of (\ref{special-relation}). Using $L_1 \equiv \sum_x P_x = \sum_x \tr W_x$, and $L_1^* = \sum_x P_x^* = \sum_x \tr W_x^{\dagger}$,
\SP{
\langle L_1^4 \rangle = &\sum_x \sum_y \sum_z \sum_w \langle P_x P_y P_z P_w \rangle \\
= &\sum_{x \ne y \ne z \ne w} \langle P_x P_y P_z P_w \rangle + 3 \sum_{x \ne y} \langle P_x^2 P_y^2 \rangle + 4 \sum_{x \ne y} \langle P_x^3 P_y \rangle + 6 \sum_{x \ne y \ne z} \langle P_x^2 P_y P_z \rangle + \sum_{x} \langle P_x^4 \rangle \, ,
\label{lhs1}
}
\SP{
\langle L_1^3 L_1^* \rangle = &\sum_x \sum_y \sum_z \sum_w \langle P_x P_y P_z P_w^* \rangle \\
= &\sum_{x \ne y \ne z \ne w} \langle P_x P_y P_z P_w^* \rangle + 3 \sum_{x \ne y} \langle P_x P_x P_y P_y^* \rangle + \sum_{x \ne y} \langle P_x^3 P_y^* \rangle + 3 \sum_{x \ne y} \langle P_x P_y P_y P_y^* \rangle \\
&+ 3 \sum_{x \ne y \ne z} \langle P_x^2 P_y P_z^* \rangle + 3 \sum_{x \ne y \ne z} \langle P_x P_y P_z P_z^* \rangle + \sum_x \langle P_x^3 P_x^* \rangle \, ,
\label{lhs2}
}
\SP{
\langle L_1^2 L_1^{* 2} \rangle = &\sum_x \sum_y \sum_z \sum_w \langle P_x P_y P_z^* P_w^* \rangle \\
= &\sum_{x \ne y \ne z \ne w} \langle P_x P_y P_z^* P_w^* \rangle + 2 \sum_{x \ne y} \langle P_x P_y P_x^* P_y^* \rangle + \sum_{x \ne y} \langle P_x P_x P_y^* P_y^* \rangle \\
&+4 \sum_{x \ne y} \langle P_x P_x P_x^* P_y^* \rangle + 4 \sum_{x \ne y \ne z} \langle P_x P_y P_x^* P_z^* \rangle + 2 \sum_{x \ne y \ne z} \langle P_x P_x P_y^* P_z^* \rangle \\
&+ \sum_x \langle P_x^2 P_x^{* 2} \rangle \, .
\label{lhs3}
}
On the r.h.s. of (\ref{special-relation}) one obtains
\SP{
\langle L_1^2 \rangle^2 = &\sum_x \sum_y \sum_z \sum_w \langle P_x P_y \rangle \langle P_z P_w \rangle \\
= &\sum_{x \ne y \ne z \ne w} \langle P_x P_y \rangle \langle P_z P_w \rangle + \sum_{x \ne y} \langle P_x^2 \rangle \langle P_y^2 \rangle + 2 \sum_{x \ne y} \langle P_x P_y \rangle^2 + 4 \sum_{x \ne y} \langle P_x^2 \rangle \langle P_x P_y \rangle \\
&+ 4 \sum_{x \ne y \ne z} \langle P_x P_y \rangle \langle P_y P_z \rangle + 2 \sum_{x \ne y \ne z} \langle P_x P_x \rangle \langle P_y P_z \rangle + \sum_x \langle P_x^2 \rangle^2 \, ,
\label{rhs1}
}
\SP{
\langle L_1^2 \rangle \langle L_1 L_1^* \rangle = &\sum_x \sum_y \sum_z \sum_w \langle P_x P_y \rangle \langle P_z P_w^* \rangle \\
= & \sum_{x \ne y \ne z \ne w} \langle P_x P_y \rangle \langle P_z P_w^* \rangle + \sum_{x \ne y} \langle P_x^2 \rangle \langle P_y P_y^* \rangle + 2 \sum_{x \ne y} \langle P_x P_y \rangle \langle P_x P_y^* \rangle \\
&+ 2 \sum_{x \ne y} \langle P_x P_y \rangle \langle P_y P_y^* \rangle + 2 \sum_{x \ne y} \langle P_x^2 \rangle \langle P_x P_y^* \rangle + 4 \sum_{x \ne y \ne z} \langle P_x P_y \rangle \langle P_x P_z^* \rangle \\
&+ \sum_{x \ne y \ne z} \langle P_x^2 \rangle \langle P_y P_z^* \rangle + \sum_{x \ne y \ne z} \langle P_x P_y \rangle \langle P_z P_z^* \rangle + \sum_x \langle P_x^2 \rangle \langle P_x P_x^* \rangle \, ,
\label{rhs2}
}
\SP{
\langle L_1 L_1^* \rangle^2 = &\sum_x \sum_y \sum_z \sum_w \langle P_x P_y^* \rangle \langle P_z P_w^* \rangle \\
= &\sum_{x \ne y \ne z \ne w} \langle P_x P_y^* \rangle \langle P_z P_w^* \rangle + \sum_{x \ne y} \langle P_x P_x^* \rangle \langle P_y P_y^* \rangle + 2 \sum_{x \ne y} \langle P_x P_y^* \rangle^2 \\
&+ 4 \sum_{x \ne y} \langle P_x P_x^* \rangle \langle P_x P_y^* \rangle + 2 \sum_{x \ne y \ne z} \langle P_x P_x^* \rangle \langle P_y P_z^* \rangle + 4 \sum_{x \ne y \ne z} \langle P_x P_y^* \rangle \langle P_x P_z^* \rangle \\
&+ \sum_x \langle P_x P_x^* \rangle^2 \, .
\label{rhs3}
}
To determine with certainty whether (\ref{special-relation}) holds in general, lattice simulations of Yang-Mills theory would be necessary. However, we can check if it is satisfied at leading order in the limit that the coupling $g^2 \rightarrow \infty$ using the effective action in (\ref{Z_YM_lambda}),
\EQ{
e^{-S_{eff}} = \prod_{\langle x y \rangle} \left[ 1 + \lambda_1 \left[ \tr(W_x) \tr(W_{y}^{\dagger}) + \tr (W_{x}^{\dagger}) \tr (W_y) \right] \right] \, ,
}
where $\lambda_1 = \left(\frac{\beta_{lat}}{2 N_c^2}\right)^{N_{\tau}}$. In this case the only expectation values which survive are those which involve $SU(3)$ color singlet contributions (see, for example, chapter $8$ in \cite{Cvitanovic:2008zz}), For general $N_c$
\EQ{
\int_{SU(N_c)} {\rm d}W~ (\tr W \tr W^{\dagger})^l (\tr W)^{N_c m} (\tr W^{\dagger})^{N_c n} \ne 0 \, ,
\label{group-int}
}
where $l$, $m$, $n = 0, 1, 2, ...$. Therefore to ${\cal O}(\lambda_1)$ the only nonzero contributions in (\ref{special-relation}) arise from
\EQ{
\langle P_x P_x^* \rangle = \int_{SU(N_c)} {\rm d}W_x~ \tr W_x \tr W_x^{\dagger} = 1 \, ,
\label{ppstar-app}
}
\EQ{
\langle P_x^2 P_x^{* 2} \rangle = \int_{SU(N_c)} {\rm d}W_x~ (\tr W_x \tr W_x^{\dagger})^2 = 2 \, ,
\label{ppstar2-app}
}
\EQ{
\langle P_x P_y P_x^{*} P_y^* \rangle = \int_{SU(N_c)} {\rm d}W_x {\rm d}W_y~ (\tr W_x \tr W_x^{\dagger}) (\tr W_y \tr W_y^{\dagger}) = 1 \, ,
}
and, if $x, y$ are nearest neighbors and $z \ne x, y$ then
\EQ{
\langle P_x P_y^{*} \rangle = \lambda_1 \int_{SU(N_c)} {\rm d}W_x {\rm d}W_y~ (\tr W_x \tr W_x^{\dagger}) (\tr W_y \tr W_y^{\dagger}) = \lambda_1 \, ,
}
\EQ{
\langle P_x P_x P_x^* P_y^{*} \rangle = \lambda_1 \int_{SU(N_c)} {\rm d}W_x {\rm d}W_y~ (\tr W_x \tr W_x^{\dagger})^2 (\tr W_y \tr W_y^{\dagger}) = 2\lambda_1 \, ,
}
\EQ{
\langle P_z P_x P_z^* P_y^{*} \rangle = \lambda_1 \int_{SU(N_c)} {\rm d}W_x {\rm d}W_y {\rm d}W_z~ (\tr W_x \tr W_x^{\dagger}) (\tr W_y \tr W_y^{\dagger}) (\tr W_z \tr W_z^{\dagger}) = \lambda_1 \, ,
}
and for the case with $N_c = 3$,
\EQ{
\langle P_x^2 P_y^{* 2} \rangle = \lambda_1 \int_{SU(3)} {\rm d}W_x {\rm d}W_y~ (\tr W_x)^3 (\tr W_y^{\dagger})^3 = \lambda_1 \, ,
}
where the results are obtained by counting the number of singlets in the decomposition: ${\bf N_c} \otimes {\bf {\bar N_c}}$ has $1$, ${\bf N_c} \otimes {\bf {\bar N_c}} \otimes {\bf N_c} \otimes {\bf {\bar N_c}}$ has $2$, ${\bf 3} \otimes {\bf 3} \otimes {\bf 3}$ and ${\bf {\bar 3}} \otimes {\bf {\bar 3}} \otimes {\bf {\bar 3}}$ both have $1$. Calculating the expectation values in (\ref{lhs1}-\ref{rhs3}) using the group integrals above one obtains
\begin{align}
\langle L_1^4 \rangle = &~0 \, , \\
\langle L_1^3 L_1^* \rangle = &~0 \, , \\
\label{h4_disc}
\langle L_1^2 L_1^{* 2} \rangle = &~2 \sum_{x \ne y} \langle P_x P_y P_x^* P_y^* \rangle + \sum_x \langle P_x^2 P_x^{* 2} \rangle + { \sum_{x \ne y} \langle P_x P_x P_y^* P_y^* \rangle} + 4 \sum_{x \ne y} \langle P_x P_x P_x^* P_y^* \rangle \nonumber\\
&~+ 4 \sum_{x \ne y \ne z} \langle P_x P_y P_x^* P_z^* \rangle + {\cal O}(\lambda_1^2) \\
= &~2 \sum_{x \ne y} + 2 \sum_x + { 18} \lambda_1 \sum_{\langle x y \rangle} + 8 \lambda_1 \sum_{x \ne \langle y z \rangle}\, + {\cal O}(\lambda_1^2) \nonumber\\
= &~2 N_s^2 + { 6} N_s \lambda_1 + 24 N_s^2 \lambda_1 + {\cal O}(\lambda_1^2) \, , \nonumber\\
\langle L_1^2 \rangle^2 = &~0 \, , \\
\langle L_1^2 \rangle \langle L_1 L_1^* \rangle = &~0 \, , \\
\langle L_1 L_1^* \rangle^2 = &\sum_{x \ne y} \langle P_x P_x^* \rangle \langle P_y P_y^* \rangle + \sum_x \langle P_x P_x^* \rangle^2 + 4 \sum_{x \ne y} \langle P_x P_x^* \rangle \langle P_x P_y^* \rangle \nonumber\\
&~+ 2 \sum_{x \ne y \ne z} \langle P_x P_x^* \rangle \langle P_y P_z^* \rangle + {\cal O}(\lambda_1^2) \\
= &~\sum_{x \ne y} + \sum_x + 8 \lambda_1 \sum_{\langle x y \rangle} + 4 \lambda_1 \sum_{x \ne \langle y z \rangle} + {\cal O}(\lambda_1^2) \nonumber\\
= &~N_s^2 + 12 N_s^2 \lambda_1 + {\cal O}(\lambda_1^2) \nonumber\, .
\end{align}
Collecting the non-zero contributions reduces (\ref{special-relation}) to
\EQ{
\langle L_1^2 L_1^{* 2} \rangle = 2 \langle L_1 L_1^* \rangle^2 \, ,
}
which is true for $\lambda_1 = 0$ via the group integrals. However, at ${\cal O}(\lambda_1)$, there is an additional contribution at ${\cal O}(V)$ (proportional to $N_s$) which arises in (\ref{h4_disc}) because $\sum_{x \ne y} \langle P_x P_x P_y^* P_y^* \rangle \ne 0$ for $SU(3)$. Notice that this quantity is zero in the limit $N_c \rightarrow \infty$ so in that case the simpler $Q \frac{Z_{YM}}{Z} = \exp[-p(p+N_f)X_1]$ structure is maintained.


\bibliography{phase}

\end{document}